\documentclass[
a4paper,
aps,
prd,
twocolumn,
tightenlines,
preprintnumbers,
nofootinbib,
showkeys,
superscriptaddress
]{revtex4-1}

\usepackage{XCharter}
\usepackage[T1]{fontenc}

\usepackage{fullpage}
\usepackage{amsfonts}
\usepackage{amsmath}
\usepackage{slashed}
\usepackage{amssymb}
\usepackage{graphicx}
\usepackage{epic}
\usepackage{eepic}
\usepackage{epsfig}
\usepackage{latexsym}
\usepackage[dvipsnames]{xcolor}
\usepackage[export]{adjustbox}
\usepackage{float}
\usepackage{multirow}
\usepackage{hyperref}
\usepackage{enumitem}
\hypersetup{colorlinks=true,citecolor=red,linkcolor=NavyBlue,urlcolor=NavyBlue}
\usepackage[caption=false]{subfig}
 
\usepackage{natbib}
\usepackage{relsize}
\usepackage[left=1.7cm,right=1.7cm,top=2.1cm,bottom=2.5cm]{geometry}
\usepackage{shorthand}

\begin{document}
\relscale{0.95}

\title{Boosting vector leptoquark searches with boosted tops}

\author{Arvind Bhaskar}
\email{arvind.bhaskar@research.iiit.ac.in}
\affiliation{Center for Computational Natural Sciences and Bioinformatics, International Institute of Information Technology, Hyderabad 500 032, India}

\author{Tanumoy Mandal}
\email{tanumoy@iisertvm.ac.in}
\affiliation{Indian Institute of Science Education and Research Thiruvananthapuram, Vithura, Kerala, 695551, India}

\author{Subhadip Mitra}
\email{subhadip.mitra@iiit.ac.in}
\affiliation{Center for Computational Natural Sciences and Bioinformatics, International Institute of Information Technology, Hyderabad 500 032, India}

\date{\today}

\begin{abstract}
\noindent
At the LHC, a TeV-scale leptoquark (LQ) that decays dominantly to a top quark ($t$) and a light charged lepton ($\ell=e,\mu$) would form a resonance system of \emph{boosted-$t$ $+$ high-$p_{\rm T}$-$\ell$}. We consider all possible vector LQ models within the Buchm\"{u}ller-R\"{u}ckl-Wyler classifications with the desired decay. We propose simple phenomenological Lagrangians that are suitable for bottom-up/experimental studies and, at the same time, can cover the relevant parameter spaces of these models. In this simplified framework, we study the pair and single production channels of vector LQs at the LHC. Interestingly, we find that, like the pair production, the cross sections of some single production processes also depend on the parameter $\kappa$ that appears in the gluon-vector LQ coupling. We adopt a search strategy of selecting events with at least one boosted hadronic top quark and exactly two high-$p_{\rm T}$ leptons of the same flavor and opposite sign. This combines events from the pair and single production processes and, therefore, can enhance the discovery potential beyond that of the pair-production-only searches. For $5\sigma$ discovery we find that vector LQs can be probed up to $2.55$ TeV for $100\%$ branching ratio in the $t\ell $ decay mode and $\mathcal{O}(1)$ new couplings at the $14$ TeV LHC with $3$ ab$^{-1}$ of integrated luminosity.
\end{abstract}

\maketitle


\section{Introduction}
\label{sec:intro}

\noindent
In the recent past, several experimental collaborations have reported some hints of lepton flavor universality violation in the heavy meson decays. Collectively, these point toward the existence of some physics beyond the Standard Model (SM) as the SM gauge interactions are flavor-blind. Intriguingly, these seem to be quite tenacious and have created a lot of excitement in the particle physics community.  
Initially, the {\sc BaBar} collaboration found two significant anomalies in the flavor-changing charged current decays of the $B$-meson via the $b\to c\tau\nu$ transition. They reported the anomalies in terms of excesses in the $R_{D^{(*)}}$ observables defined as the ratios of branching 
ratios (BRs) to reduce some systematic and hadronic uncertainties~\cite{Lees:2012xj,Lees:2013uzd}. 
Since then, the excesses have survived the later measurements by the LHCb~\cite{Aaij:2015yra,Aaij:2017uff,Aaij:2017deq} and Belle~\cite{Huschle:2015rga,Hirose:2016wfn,Hirose:2017dxl,Abdesselam:2019dgh} collaborations. The statistical average of these two observables  obtained in the $R_D$-$R_{D^{*}}$ plane by the HFLAV Group puts the anomalies away from the corresponding SM predictions~\cite{Bigi:2016mdz,Bernlochner:2017jka,Bigi:2017jbd,Jaiswal:2017rve} by a combined significance of $\sim 3.1\sg$.
The LHCb Collaboration has also observed downward deviations of about $2.5\sg$~\cite{Aaij:2014pli,Aaij:2014ora,Aaij:2015oid,Aaij:2017vbb,Aaij:2019wad} from the
SM predictions~\cite{Hiller:2003js,Bordone:2016gaq} in the flavor-changing neutral current transition $b\to s\mu\mu$ measured in terms of the
$R_{K^{(*)}}$ observables. Similarly, an excess of about $\sim 2\sg$ was found in another observable, $R_{J/\psi}$~\cite{Aaij:2017tyk}. In addition, a long-standing discrepancy of about $\sim 3.5\sg$  exists in the muon anomalous magnetic moment measurement~\cite{Bennett:2006fi}.

It is known that TeV-scale
leptoquarks (LQs) are good candidates to address the flavor anomalies. Moreover, 
their phenomenology has been explored in various other contexts as well~\cite{Sakaki:2013bfa,
Mohanta:2013lsa,
Sahoo:2015wya,
Dey:2015eaa,
Mandal:2015lca,
Freytsis:2015qca,
Sahoo:2015qha,
Sahoo:2015fla,
Aydemir:2015oob,
Sahoo:2015pzk,
Aydemir:2016qqj,
Das:2016vkr,
Sahoo:2016nvx,
Becirevic:2016oho,
Bandyopadhyay:2016oif,
Sahoo:2016pet,
Faroughy:2016osc,
Hiller:2016kry,
Bhattacharya:2016mcc,
Duraisamy:2016gsd,
Das:2017kkm,
Assad:2017iib,
Dey:2017ede,
Biswas:2018jun,
Bandyopadhyay:2018syt,
Aydemir:2018cbb,
Sahoo:2018ffv,
Kumar:2018kmr,
Crivellin:2018yvo,
Mandal:2018qpg,
Biswas:2018snp,
Mandal:2018czf,
Angelescu:2018tyl,
Balaji:2018zna,
Mandal:2018kau,
Alvarez:2018jfb,
Biswas:2018iak,
Iguro:2018vqb,
Aebischer:2018acj,
Roy:2018nwc,
Fornal:2018dqn,
Bar-Shalom:2018ure,
Kim:2018oih,
Alves:2018krf,
Baker:2019sli,
Aydemir:2019ynb,
Cornella:2019hct,
Zhang:2019jwp,
Mandal:2019gff,
Hou:2019wiu,
Coy:2019rfr,
Allanach:2019zfr,
Padhan:2019dcp,
Bhaskar:2020kdr,
Bandyopadhyay:2020klr}. 
LQs are color-triplet bosons [either scalars (sLQs) or vectors (vLQs)] predicted by many beyond-the-SM theories~\cite{Pati:1974yy,Georgi:1974sy,Schrempp:1984nj,Barbier:2004ez,Kohda:2012sr}.
They have fractional electric charges and carry both lepton and baryon numbers. In general, a LQ can couple to a quark and a lepton of either the same or different generations. The flavor anomalies suggest that LQs couple more strongly to the third-generation fermions than the other two.
Cross-generational couplings of LQs could generate flavor-changing neutral currents; those involving the first and second generations are tightly constrained. However,
bounds are relatively weaker when a fermion of the third-generation is involved. 

The search for LQs is an important research program at the LHC.
Usually, the LHC searches are done for LQs that couple to quarks and leptons of the same generation and are labeled accordingly. For example, pair production of a scalar LQ that decays to a top quark and a tau lepton (or a bottom quark and a tau lepton or neutrino), i.e, a third-generation LQ, has been extensively analyzed by both the 
ATLAS~\cite{Aaboud:2019jcc,Aaboud:2019bye} and CMS~\cite{Sirunyan:2018nkj,Sirunyan:2018kzh} collaborations. 
Altogether, the current bound on the third-generation LQ is roughly about a TeV (this, of course, depends on various assumptions and we refer the reader to the actual papers for details). 
However, the flavor-motivated LQ models with sizable cross-generational couplings would have exotic signatures and require different search strategies. 
Of late, the nonstandard decay modes of LQs have started to gain attention; the CMS Collaboration published their first results on the pair production searches of LQs in the $tt\mu\mu$
channel~\cite{Sirunyan:2018ruf}. Based on the $13$ TeV data, they performed a prospect study for the pair production of sLQs in the  $tt\mu\mu$
channel at the high-luminosity LHC (HL-LHC)~\cite{CMS:2018yke}.

In Ref~\cite{Chandak:2019iwj}, we investigated the HL-LHC prospects of sLQs that couple dominantly to the top quark in some detail. There, we focused on charge $1/3$ and $5/3$ sLQs that decay to a top quark and a charged lepton. Even though we considered only third-generation quarks, interestingly, we found that in some scenarios single production can improve their prospects significantly.\footnote{This is interesting because, when a LQ (or any other particle) mostly couples  to the third-generation quarks, we generally expect their single production to be ignorable as the bottom (top) quark density in the proton is too small (nonexistent) to play any significant role at the LHC.} In this paper, we present a similar follow-up study for vLQs. Here, too, we concentrate on a specific subset of possible vLQs that dominantly couple with a top quark and can decay to a top quark and a light charged lepton ($e$ or $\m$) with a substantial BR. Since an analysis of the pair production of vLQs that decay to a top quark and a neutrino at the LHC is available in Ref.~\cite{Vignaroli:2018lpq}, in this paper we do not analyze this channel again. Instead, we present a set of simplified models that covers all of the possibilities of a vLQ decaying to a top quark and any lepton. These are suitable for experimental analysis. We also demonstrate how they are related to the known models of vLQs~\cite{Buchmuller:1986zs,Blumlein:1994qd,Blumlein:1996qp,Dorsner:2016wpm}.

Our main motivation for considering this specific type of vLQs is to investigate their collider discovery/exclusion potential by making use of the boosted top signature coming from the LQ decay.
They form an exotic resonance system with a boosted top and a high-$p_{\rm T}$ lepton and provide a novel way to search for these models at the LHC.
Various flavor anomalies suggest that cross-generational Yukawa-type LQ couplings with tops and leptons might be large. A large coupling makes various single production channels important, especially in the high-mass region. For example, the relevance of the process $gg\to{\rm LQ}+t\mu$ was pointed out in Ref.~\cite{Camargo-Molina:2018cwu}. In our analysis, we adopt the same search strategy as the one we proposed for the sLQs~\cite{Chandak:2019iwj}. We identify our signal by selecting at least one boosted hadronic top and exactly two high-$p_{\rm T}$ leptons and use the highest-$p_{\rm T}$ top
(if there is more than one top) and one of the selected leptons to reconstruct a heavy system, i.e., the LQ.
As we have demonstrated before~\cite{Mandal:2012rx,Mandal:2015vfa,Mandal:2016csb,Chandak:2019iwj}, such a selection strategy  combines pair and single production events and increases the LHC reach. 
Although pair production is suitable for probing the low-mass region, single production takes over when the LQ becomes heavy.
Compared to the sLQs, the pair production cross sections for vLQs are relatively bigger 
and hence the current mass limits obtained for pair production are generally higher for the vLQs than for the sLQs. In the case of vLQs, the importance of single production becomes visible for relatively higher mass compared to sLQs. 
We shall see that the discovery prospects of the vLQs at the HL-LHC is significantly improved if the new couplings controlling single production are of order unity.

Before we proceed further, we note that since this paper is a follow-up to Ref.~\cite{Chandak:2019iwj}, we shall frequently refer to that paper and omit some details that are common while ensuring that our presentation is self-contained. The rest of the paper is organized as follows. In Sec.~\ref{sec:model}, we describe the vLQ models and introduce simplified models suitable for experimental analysis. In Sec.~\ref{sec:pheno}, we discuss
the LHC phenomenology and illustrate our search strategy, and then we present our estimations in Sec.~\ref{sec:dispot}. Finally, we summarize
and conclude in Sec.~\ref{sec:End}.


\section{Vector Leptoquark Models}\label{sec:model}

\noindent
To conserve electromagnetic charge, vLQs that decay to a top-lepton pair would have either electric charge equal to $\pm$1/3 or $\pm$5/3 (if the lepton is a charged one) or $2/3$ (if lepton is a neutrino). This means that among the vLQs listed in Refs.~\cite{Buchmuller:1986zs,Blumlein:1994qd,Blumlein:1996qp,Dorsner:2016wpm}, the weak singlets $U_1$ and $\tilde{U}_{1}$, doublets $V_{2}$ and $\tilde V_2$ and triplet $U_3$ would qualify for our study. 
Below, we display the relevant terms in the interaction Lagrangians following the notation of Ref.~\cite{Dorsner:2016wpm}. To avoid proton decay constraints, we ignore the diquark operators. \bigskip

\noindent
$\blacksquare\quad$\underline{$\tilde{U}_{1}$ = (${\mathbf{3}}$,$\mathbf{1}$,$5/3$):}\quad
The electric charge of $\tilde{U}_{1}$ is $5/3$. Hence, it couples exclusively with the right-handed leptons:
\be \label{eq:LagU1t}
\mc L \supset \tilde{x}_{1~ij}^{RR}\bar{u}_{R}^{i}\gamma^{\mu}\tilde{U}_{1,\m} \ell_{R}^{j} + \textrm{H.c.},
\ee
where $u_{R}$ and $\ell_{R}$ are a SM right-handed up-type quark and a charged lepton, respectively, and $ i,j = \{1,2,3\}$ are the generation indices. The color indices are suppressed. 
For our purpose, we consider only those terms that would connect a vLQ to a third-generation quark and ignore the rest,
\begin{equation} \label{eq:LagU1tp}
\mathcal{L} \supset \tilde{x}_{1~3j}^{RR}~\bar{t}_{R} \lt(\gm\cdot\tilde{U}_{1}\rt) \ell^j_{R} +  \textrm{H.c.}
\end{equation}

\noindent
$\blacksquare\quad$\underline{$U_{1}$ = ($\mathbf{3}$,$\mathbf{1}$,$2/3$):}\quad
The necessary interaction terms for the charge-$2/3$ $U_1$ can be written as
\ba 
\label{eq:LagU1}
\mathcal{L} \supset x_{1\ ij}^{LL}~\bar{Q}_{L}^{i}\gamma^{\mu}U_{1,\mu}L_{L}^{j}  + x_{1\ ij}^{RR}~\bar{d}_{R}^{i}\gamma^{\mu}U_{1,\mu}\ell_{R}^{j}+ \textrm{H.c.},
\ea
where $Q_L$, $L_L$, and $d_R$ are the SM left-handed quark doublet, lepton doublet, and a down-type right-handed quark, respectively.
The $i=3$ terms can be written explicitly as
\begin{align}
\label{eq:LagU1p}
\mathcal{L} &\supset x_{1\ 3j}^{LL}\left\{\bar{t}_{L}\lt(\gamma\cdot U_{1}\rt)\n_{L}^{j} + \bar{b}_{L}\lt(\gamma\cdot U_{1}\rt)\ell_L^{j}\rt\} \nn\\
&+x_{1\ 3j}^{RR}~\bar{b}_{R}\lt(\gamma\cdot U_{1}\rt)\ell_R^{j}+\textrm{H.c.}
\end{align}

\noindent
$\blacksquare\quad$\underline{$V_{2}$ = ($\bar{\mathbf{3}}$,$\mathbf{2}$,$5/6$):}\quad
For $V_{2}$, the Lagrangian is as follows:
\begin{align}
\label{eq:LagV2}
\mathcal{L} &\supset x_{2\ ij}^{RL}~\bar{d}_{R}^{Ci}\gamma^{\mu} V_{2,\mu}^{a} \epsilon^{ab} L_{L}^{jb}\nn\\
&+ x_{2\ ij}^{LR}~\bar{Q}_{L}^{Ci,a}\gamma^{\mu}\epsilon^{ab} V_{2,\mu}^{b} \ell_{R}^{j}+
 \textrm{H.c.}
\end{align}
The superscript $C$ denotes  charge conjugation. 
Expanding the Lagrangian we get,
\begin{align}
\mathcal{L} &\supset 
-( x_{2}^{RL}\mathbf U)_{ij}\bar{d}_{R}^{Ci}\gamma^{\mu} V_{2,\mu}^{1/3} \n_{L}^{j} 
+x_{2\ ij}^{RL}~\bar{d}_{R}^{Ci}\gamma^{\mu} V_{2,\mu}^{4/3}\ell_{L}^{j}\nn\\
&+~(\mathbf V^{T}x_{2}^{LR})_{ij}\bar{u}_{L}^{Ci}\gamma^{\mu} V_{2,\mu}^{1/3} \ell_{R}^{j} - 
x_{2\ ij}^{LR}\bar{d}_{L}^{Ci}\gamma^{\mu} V_{2,\mu}^{4/3}\ell_{R}^{j} + \textrm{H.c.},
\end{align}
where $\mathbf U$ and $\mathbf V$ represent the Pontecorvo-Maki-Nakagawa-Sakata (PMNS) neutrino mixing matrix and the Cabibbo-Kobayashi-Maskawa (CKM) quark mixing matrix, respectively. We assume $\mathbf U$ to be unity, as the LHC is blind to the flavor of the neutrinos. Similarly, since the small off-diagonal terms of the CKM matrix play a negligible role at the LHC, we assume a diagonal CKM matrix for simplicity. Hence, the terms relevant for our analysis are
\begin{align}
\mathcal{L} &\supset  
-x_{2\ 3j}^{RL}\bar{b}_{R}^{C}\left\{\lt(\gamma\cdot V_{2}^{1/3}\rt)\n^j_L
- \lt(\gamma\cdot V_{2}^{4/3}\rt)\ell_{L}^{j}\rt\}\nn\\
&+x_{2\ 3j}^{LR}\left\{\bar{t}_{L}^{C}\lt(\gamma\cdot V_{2}^{1/3}\rt)
- \bar{b}_{L}^{C}\lt(\gamma\cdot V_{2}^{4/3}\rt)\rt\}\ell_{R}^{j} + \textrm{H.c.} \label{eq:LagV2p}
\end{align}

\noindent
$\blacksquare\quad$\underline{$\tilde V_{2}$ = ($\bar{\mathbf{3}}$,$\mathbf{2}$,$-1/6$):}\quad
For $\tilde V_{2}$, the Lagrangian becomes
\begin{align}
\label{eq:LagV2t}
\mathcal{L} \supset&\ \tilde{x}^{RL}_{2\ ij}\bar{u}_{R}^{C\,i} \gamma^\mu \tilde{V}^{b}_{2,\mu} \epsilon^{ab}L_{L}^{j,a} +\textrm{H.c.}
\end{align}
Expanding it, we get
\begin{align}
\mathcal{L} &\supset -\tilde{x}^{RL}_{2\ ij}\bar{u}_{R}^{C\,i} \gamma^\mu \tilde{V}^{1/3}_{2,\mu} \ell_{L}^{j} +(\tilde{x}^{RL}_2\mathbf  U)_{ij}\bar{u}_{R}^{C\,i} \gamma^\mu \tilde{V}^{-2/3}_{2,\mu} \nu_{L}^{j}\nn\\
&+\textrm{H.c.}
\end{align}
The terms with the third-generation quarks are
\begin{align}\label{eq:LagV2tp}
\mathcal{L} \supset&\ \tilde{x}^{RL}_{2\ 3j}\bar{t}_{R}^{C}\lt\{ -\lt(\gamma\cdot \tilde{V}^{1/3}_{2}\rt) \ell_{L}^{j} + \lt(\gamma\cdot \tilde{V}^{-2/3}_{2}\rt) \nu_{L}^{j}\rt\}+\textrm{H.c.}
\end{align}

\begin{table*}[t]
\begin{tabular*}{\textwidth}{l @{\extracolsep{\fill}}c @{\extracolsep{\fill}}ccc  @{\extracolsep{\fill}}cc @{\extracolsep{\fill}}cc}\hline
&&\multicolumn{3}{c}{Simplified models [Eqs.~\eqref{eq:simplelag1} -- \eqref{eq:simplelag5}]}&\multicolumn{2}{c}{LQ models [Eqs.~\eqref{eq:LagU1t} -- \eqref{eq:LagU3p}]}&&\\\cline{3-5}\cline{6-7}
\begin{tabular}[c]{l}Benchmark \\ scenario\end{tabular} & \begin{tabular}[c]{c}Possible \\ charge(s)\end{tabular} & \begin{tabular}[c]{c}Type of  \\ LQ\end{tabular} & \begin{tabular}[c]{c}Nonzero \\ couplings\\ equal to $\lm$\end{tabular} & \begin{tabular}[c]{c}Charged\\lepton \\ chirality \\fraction\end{tabular} & \begin{tabular}[c]{c}Type of  \\ LQ\end{tabular} &\begin{tabular}[c]{c}Nonzero \\ coupling\\ equal to $\lm$\end{tabular} & \begin{tabular}[c]{c}Decay \\ mode(s)\end{tabular}& \begin{tabular}[c]{c}Branching \\ ratios(s)\\$\{\bt,1-\bt\}$\end{tabular}\\\hline\hline\
\multirow{3}{*}{LC}      & $1/3$   &   $\chi_1$  &   $\Lm_\ell$        & $\eta_L=1$      &  $\displaystyle\tilde V_2^{1/3}$        & $\tilde x_{2\ 3j}^{RL}$           & $t\ell$ & \multirow{3}{*}{$\{100\%,0\}$}\\
        & $2/3$  &   $\chi_2$  &   $\bar\Lm_\n$      & ---                        & $\displaystyle\lt(\tilde V_2^{-2/3}\rt)^\dag$       & $\displaystyle\lt(\tilde x_{2\ 3j}^{RL}\rt)^*$           & $t\n$   &\\
        & $5/3$   &   $\chi_5$  &   $\tilde\Lm_\ell$  & $\eta_L=1$   &     $\displaystyle U_3^{5/3}$        & $\sqrt{2}\ x_{3\ 3j}^{LL}$ & $t\ell$ & \\
\hline
LCSS*  	&\multirow{2}{*}{$2/3$}	   & \multirow{2}{*}{$\chi_2$}	& $\bar\Lambda_{\ell}=\bar\Lambda_{\nu}$        & \multirow{2}{*}{$\eta_{L}=1$}&	 $\displaystyle U_1$     					&$x_{1\ 3j}^{LL}$  & \multirow{2}{*}{\{$t\n$, $b\ell$\}}	& \multirow{2}{*}{\{$50\%$, $50\%$\}}  \\
LCOS  	&	    &	& $\bar\Lambda_{\ell}=-\bar\Lambda_{\nu}$        && $\displaystyle U_3^{2/3}$     					&$-x_{3\ 3j}^{LL}$  &  	&  \\
\hline
\multirow{2}{*}{RC}        & $1/3$   &   $\chi_1$  &   $\Lm_\ell$        & \multirow{2}{*}{$\eta_R=1$}       & $\displaystyle V_2^{1/3}$               & $x_{2\ 3j}^{LR}$                  & \multirow{2}{*}{$t\ell$}  & \multirow{2}{*}{$\{100\%,0\}$}\\
        & $5/3$   &   $\chi_5$  &   $\tilde\Lm_\ell$       &    & $\displaystyle\tilde U_1$              & $\tilde x_{1\ 3j}^{RR}$           & &\\
\hline
RLCSS*  	&\multirow{2}{*}{$1/3$}	   & \multirow{2}{*}{$\chi_1$}	& $\Lambda_{\ell}=\Lambda_{\nu}$                & \multirow{2}{*}{$\eta_{R}=1$}	& $\displaystyle V_2^{1/3}$     					&$x_{2\ 3j}^{LR}=-x_{2\ 3j}^{RL}$  & \multirow{2}{*}{\{$t\ell$, $b\n$\}} 	& \multirow{2}{*}{\{$50\%$, $50\%$\}}  \\
RLCOS*  	&	   & 	& $\Lambda_{\ell}=-\Lambda_{\nu}$                & 	& $\displaystyle V_2^{1/3}$     					&$x_{2\ 3j}^{LR}=x_{2\ 3j}^{RL}$  &&\\\hline  
\end{tabular*}
\caption{Summary of the nine benchmark scenarios considered. The branching ratio for a $\chi$ to decay to a top quark, $\bt$ is fixed for all models [Eqs.~\eqref{eq:LagU1t} -- \eqref{eq:LagU3p}], except for $U_1$ in the LCSS scenario $(\bt\leq 50\%)$ and $V_2^{1/3}$ in the RLCSS/OS scenarios where $0\leq\bt<100\%$ (for $\bt=100\%$, these two scenarios become the same as the RC scenario). The exceptional scenarios are marked by an asterisk. Here, $\lm$ is a generic free coupling parameter.  For simplicity, we have chosen only this one coupling to control all of the nonzero new couplings in every benchmark. This essentially means also choosing $\bt$ to be $50\%$ in the exceptional scenarios.}\label{tab:benchmark}
\end{table*}

\noindent
$\blacksquare\quad$\underline{$U_{3}$ = ($\mathbf{3}$,$\mathbf{3}$,$2/3$):}\quad
The necessary interaction terms for the triplet $U_3$ are
\ba 
\label{eq:LagU3}
\mathcal{L} \supset x_{3\ ij}^{LL}\bar{Q}_{L}^{i,a}\gamma^{\mu}\lt(\ta^k U_{3,\mu}^k\rt)^{ab}L_{L}^{j,b}  + \textrm{H.c.},
\ea
where $\tau^k$ denotes the Pauli matrices. This can be expanded as
\begin{align}
\nonumber
\mathcal{L} &\supset -x^{LL}_{3\ ij}\bar{d}_{L}^{i} \gamma^\mu U^{2/3}_{\mu}  \ell_{L}^{j}+(\mathbf V x^{LL}_3 \mathbf U)_{ij}\bar{u}_{L}^{i} \gamma^\mu U^{2/3}_{\mu}  \nu_{L}^{j}\\
&+\sqrt{2}(x^{LL}_3 \mathbf U)_{ij}\bar{d}_{L}^{i} \gamma^\mu U^{-1/3}_{\mu}  \nu_{L}^{j}+\sqrt{2}(\mathbf V x^{LL}_3)_{ij}\bar{u}_{L}^{i} \gamma^\mu U^{5/3}_{\mu}  \ell_{L}^{j}\nn\\
&+\textrm{H.c.}
\end{align}
The terms for the third-generation quarks can be written explicitly as,
\begin{align}
\mathcal{L} &\supset  x^{LL}_{3\ 3j}\Big\{-\bar{b}_{L} \lt(\gamma\cdot U^{2/3}_3\rt) \ell_{L}^{j}+\bar{t}_{L} \lt(\gamma\cdot U^{2/3}_3\rt) \nu_{L}^{j}\nn\\
&+\sqrt{2}\ \bar{b}_{L} \lt(\gamma\cdot U^{-1/3}_3\rt) \nu_{L}^{j}+\sqrt{2}\ \bar{t}_{L} \lt(\gamma\cdot U^{5/3}_3\rt)  \ell_{L}^{j}\Big\}\nn\\
&+\textrm{H.c.}\label{eq:LagU3p}
\end{align}

\subsection{Simplified model and benchmark scenarios}\label{subsec:benchmark}
\noindent The above models can be simplified into the following phenomenological Lagrangians:
\ba 
\label{eq:simplelag1}
\mathcal{L} &\supset& \Lambda_{\ell}\bigg\{\sqrt{\eta_{R}}\ \bar{t}_{L}^{C}\lt(\gamma\cdot\chi_1\rt)\ell_{R} +   \sqrt{\eta_{L}}\ \bar{t}_{R}^{C}\lt(\gamma\cdot\chi_1\rt)\ell_{L}\bigg\}\nn\\
&&+~\Lambda_{\n}\ \bar{b}_{R}^{C}\lt(\gamma\cdot\chi_1\rt)\n_{L}+{\rm H.c.},\\
\label{eq:simplelag2} 
\mathcal{L} &\supset& 
\bar \Lambda_{\ell}\bigg\{\ep_R\ \sqrt{\eta_{R}}\ \bar{b}_{R}\lt(\gamma\cdot\chi_2\rt)\ell_{R} + \sqrt{\eta_{L}}\ \bar{b}_{L}\lt(\gamma\cdot\chi_2\rt)\ell_{L}\bigg\}\nn\\
&&+~\bar\Lambda_{\n}\ \bar{t}_{L}\lt(\gamma\cdot\chi_2\rt)\n_L +{\rm H.c.},\\
\label{eq:simplelag5}
\mathcal{L} &\supset& \tilde{\Lambda}_{\ell}\bigg\{\sqrt{\eta_{R}}\  \bar{t}_{R}\lt(\gamma\cdot\chi_5\rt)\ell_{R}+\sqrt{\eta_{L}}\ \bar{t}_{L}\lt(\gamma\cdot\chi_5\rt)\ell_{L}\bigg\}\nn\\
&&+~{\rm H.c.},
\ea
where we have suppressed the lepton generation index. We denote a generic charge $\pm n/3$ vLQ by $\chi_{n}$. Here, $\eta_{L}$ and $\eta_{R} = 1 -\eta_{L}$ are the charged lepton chirality fractions ~\cite{Mandal:2015vfa,Chandak:2019iwj}. In Eq.~\eqref{eq:simplelag2}, we have introduced a sign term $\ep_R = \pm1$ to incorporate a possible relative sign between the left-handed and right-handed terms [see Eq.~\eqref{eq:LagV2p}]. We shall consider only real couplings in our analysis for simplicity. 

\begin{figure*}[]
\captionsetup[subfigure]{labelformat=empty}
\subfloat[(a)]{\includegraphics[height=3cm,width=4.1cm]{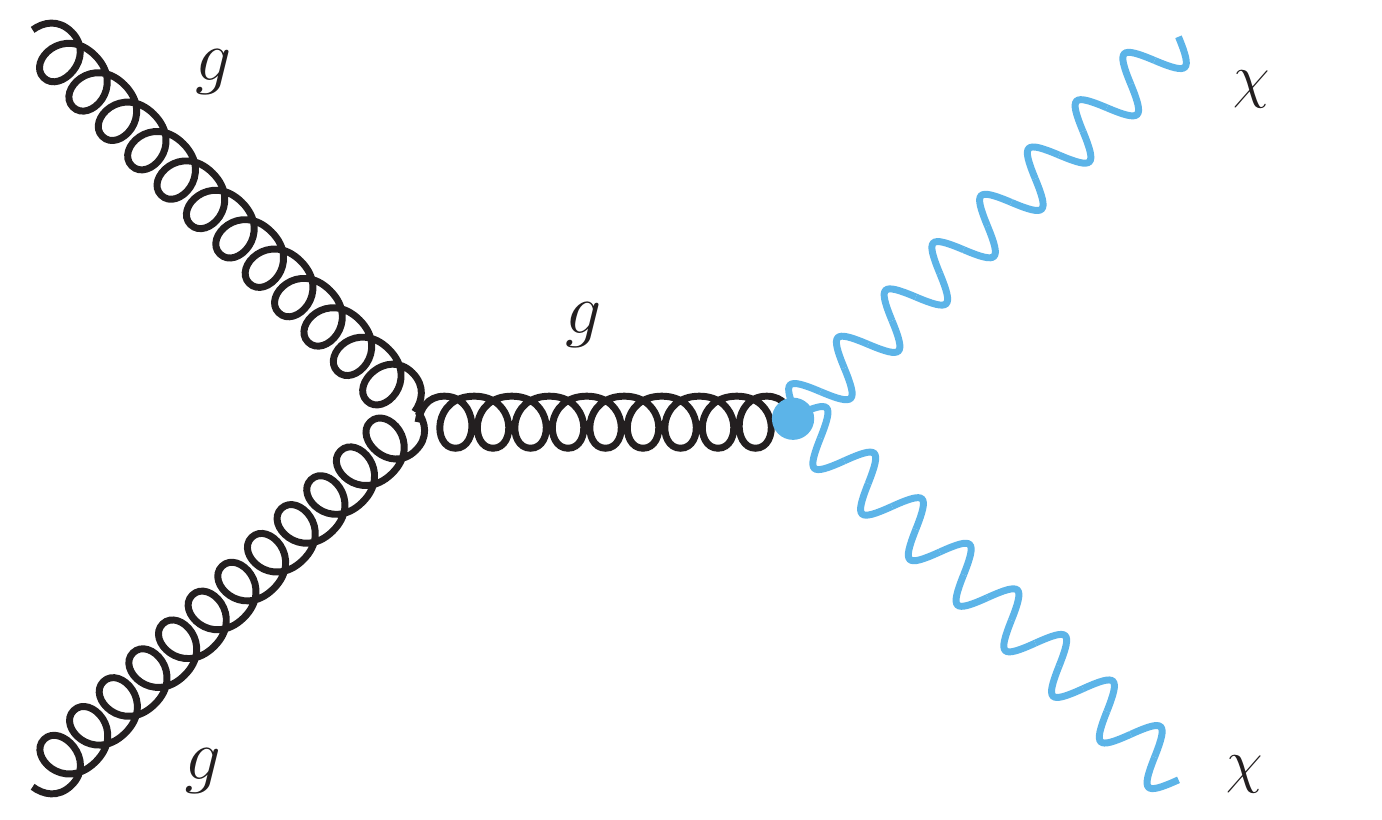}\label{fig:feyndiag01}}
\subfloat[(b)]{\includegraphics[height=3cm,width=4.1cm]{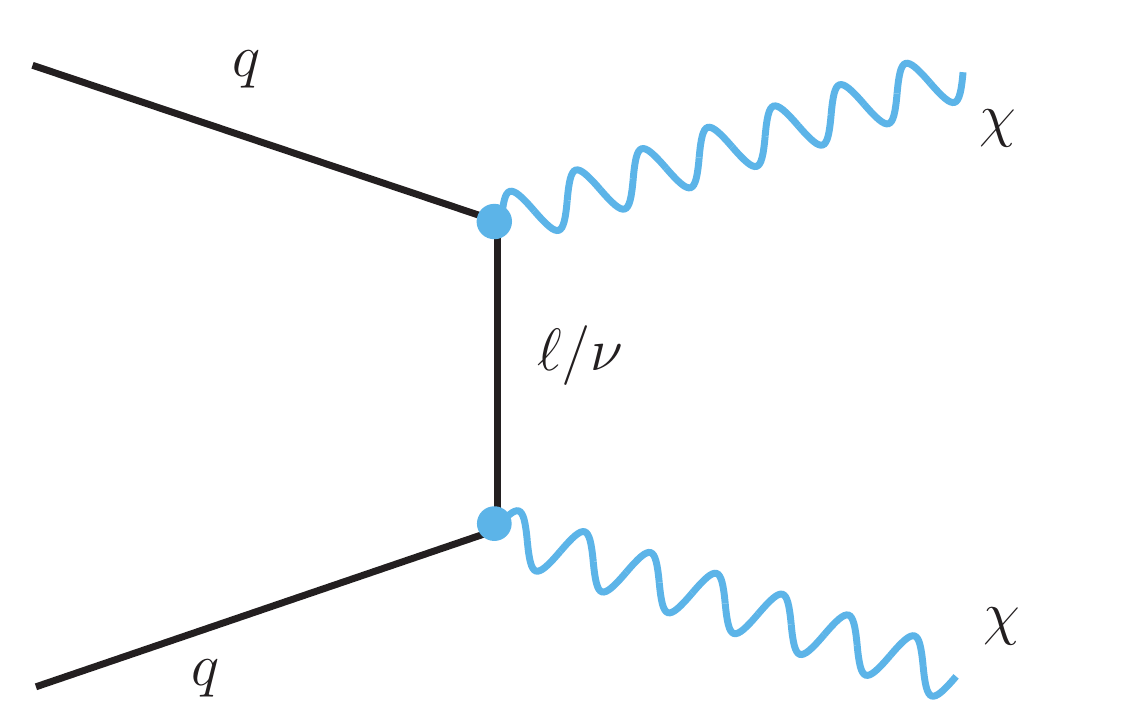}\label{fig:feyndiag11}}
\subfloat[(c)]{\includegraphics[height=3cm,width=4.1cm]{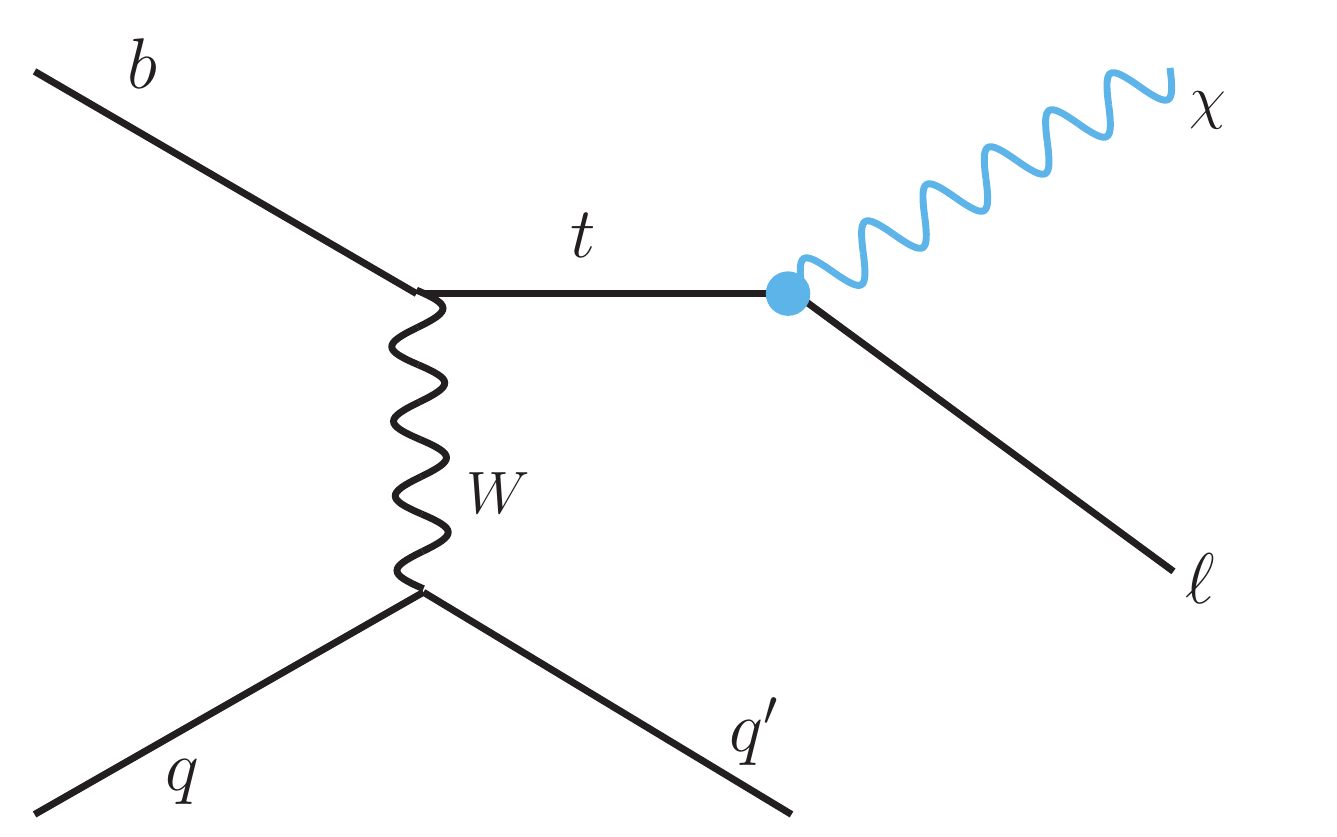}\label{fig:feyndiag02}}
\subfloat[(d)]{\includegraphics[height=3cm,width=4.1cm]{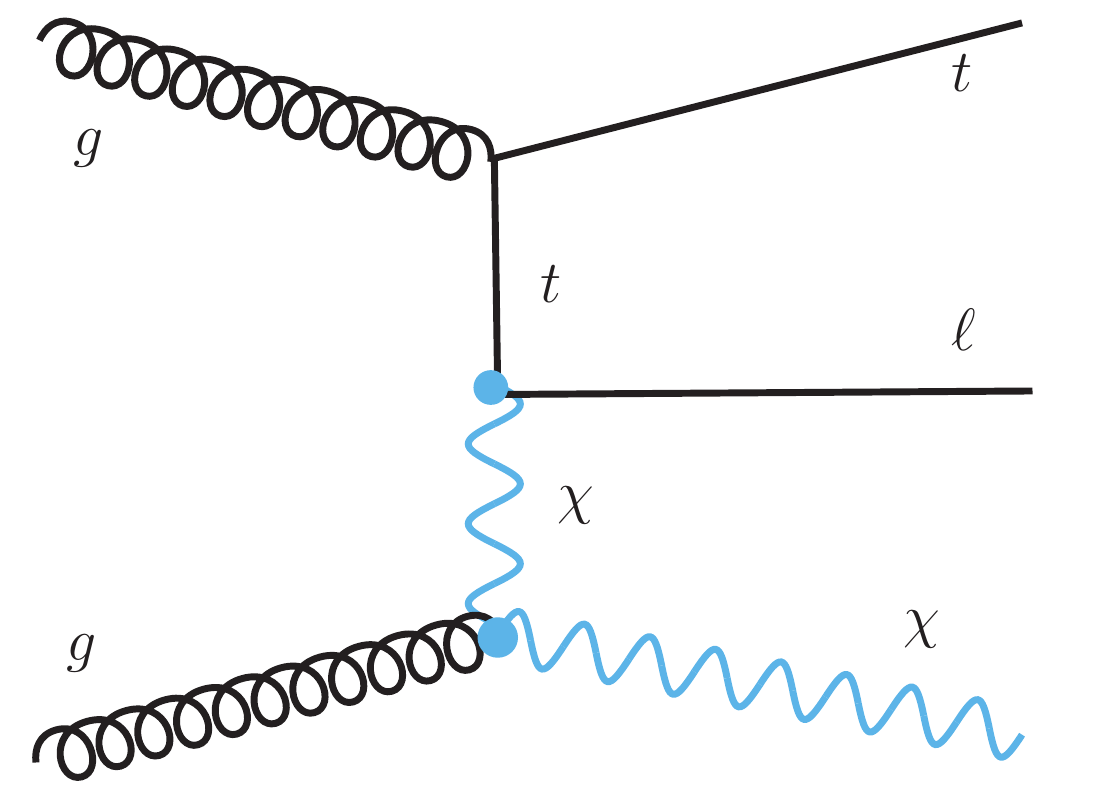}\label{fig:feyndiag12}}
\caption{\label{fig:feyndiag} Representative Feynman diagrams for the LQ production at the LHC. Panels (a) and (b) show pair production processes and panels (c) and (d) show single production processes.}
\end{figure*}

As we did for the sLQs~\cite{Chandak:2019iwj}, here \tcr{too} we identify some benchmark scenarios with the simplified models (see Table~\ref{tab:benchmark}). 
Each scenario corresponds to one of the realizable models described above [see Eqs.~\eqref{eq:LagU1t} -- \eqref{eq:LagU3p}]. Here, we have ignored any possible mixing among the vLQs. 
The BR for a $\chi$ to decay to a top quark, $\bt$, is fixed in all models [Eqs.~\eqref{eq:LagU1t} -- \eqref{eq:LagU3p}], except for two cases that we shall describe shortly.
For simplicity, we choose only one free coupling $\lm$ parametrizing the nonzero new couplings in every benchmark scenario. (See the fourth and seventh columns of Table.~\ref{tab:benchmark}. By doing this, we are essentially also choosing $\bt$ to be $50\%$ in the free $\bt$ scenarios.)

\begin{itemize}
\item 
In the Left Coupling (LC) scenario, a $\chi$ can directly couple with left-handed leptons. We set $\lm$ equal to $\Lm_\ell$ (for $\chi_1$),
$\tilde\Lm_\ell$ (for $\chi_5$) or $\bar\Lm_\n$ (for $\chi_2$) and put all other couplings to zero. For $\chi_1$ and $\chi_5$ we set $\eta_L=1$. Here ,
$\chi_1$ represents a $\tilde V_2^{1/3}$ with $\Lm_\ell=\tilde x_{2\ 3j}^{RL}$, $\chi_5$ represents a $U_3^{5/3}$ with 
$\tilde \Lm=\sqrt{2}\ x_{3\ 3j}^{LL}$ and $\chi_2$ represents an anti-$\tilde V_2^{-2/3}$ with 
$\bar\Lm_\n=\displaystyle\lt(\tilde x_{2\ 3j}^{RL}\rt)^*$. 
In this scenario, a $\chi_1$ or $\chi_5$ decays to $t\ell$ pairs and a $\chi_2$ decays to $t\n$ pairs all of the time. 

\item 
If a $\chi_2$ is of $U_1$ or $U^{2/3}_3$ type, it can also decay to a $b\ell$ pair. Hence, it is possible that a $\chi_2$ couples with 
left-handed leptons but the BR  for the decay $\chi_2\to t\n$ $(\bt)$ is $50\%$.
Such possibilities are captured in the Left Couplings with the
Same Sign (LCSS) or the Left Couplings with Opposite Signs (LCOS) scenarios. The difference between these two comes from the different relative signs
between the $\chi_2b\ell$ and $\chi_2t\n$ couplings. In LCSS, the $\chi_2$ behaves as a $U_1$ with 
$\bar\Lambda_{\ell}=\bar\Lambda_{\nu}=x_{1\ 3j}^{LL}$, whereas in LCOS it behaves as a $U_3^{2/3}$ with 
$\bar\Lambda_{\ell}=-\bar\Lambda_{\nu}=-x_{3\ 3j}^{LL}$. It is important to note that in the $U_1$ case, it is possible to have $\bt < 50\%$ if we consider a nonzero $x_{1\ 3j}^{RR}$. This can be seen from Eq.~\eqref{eq:LagU1p}.
Unlike the sLQ case~\cite{Chandak:2019iwj}, the LCSS and LCOS scenarios for the vLQ yield the same single production cross section as there is no interference among the
contributing diagrams.

\item
The Right Coupling (RC) scenario, where a LQ couples only to right-handed charged leptons, is exclusive to $\chi_1$ and $\chi_5$. Like the LC scenario, here we have $\bt=100\%$. In this case a $\chi_1$ behaves as a $V_2^{1/3}$ with $\Lm_\ell = x_{2\ 3j}^{LR}$ and a $\chi_5$ behaves as a 
$\tilde U_1$ with $\tilde \Lm_\ell = x_{1\ 3j}^{RR}$.

\item
Unlike the sLQ $\phi_1$ (see Ref.~\cite{Chandak:2019iwj}), the $\chi_1$ type vLQs (if it is $V_2^{1/3}$) can decay to both  $t\ell$ and $b\nu$ pairs, provided 
$\Lm_{\ell}$ and $\Lm_{\nu}$ are both nonzero. We design two scenarios, namely, Right (lepton) Left (neutrino) Couplings with the Same Sign (RLCSS) where $\chi_1\equiv V_2^{1/3}$ with $\Lm_\ell=\Lm_\n=x_{2\ 3j}^{LR}=x_{2\ 3j}^{RL}$, and 
Right (lepton) Left (neutrino) Couplings with Opposite Signs (RLCOS) where $\chi_1\equiv V_2^{1/3}$ with
$\Lm_\ell=-\Lm_\n=x_{2\ 3j}^{LR}=x_{2\ 3j}^{RL}$. In these two scenarios, $\bt$ can be anything between $0$ and $100\%$ as both involve two independent couplings [$x_{2\ 3j}^{LR}$ and $x_{2\ 3j}^{RL}$, see Eq.~\eqref{eq:LagV2p}]. However, we consider only $\bt=50\%$ for these benchmarks. We introduce these two benchmarks for completeness, though for our purpose these two are equivalent. As there is no interference contribution sensitive to this sign flip, all of the production processes would have the same cross sections in both scenarios.
\end{itemize}

Before we move on, we note that the kinetic terms for a vector  leptoquark contains a free parameter, usually denoted as $\kp$~\cite{Dorsner:2016wpm},
\begin{align}
\mc L \supset -\frac12\chi^\dag_{\m\n}\chi^{\m\n} + M^2_\chi\ \chi^\dag_\m\chi^\m -ig_s\kp \ \chi^\dag_\m T^a\chi_\n\ G^{a\,\m\n},\label{eq:lkin}
\end{align}
where $\chi_{\m\n}$ stands for the field-strength tensor of $\chi$. This parameter $\kp$ can change the pair and (interestingly) some single production cross sections through the 
modification of the $\chi \chi g$ vertex.\footnote{Similar modifications are also possible for other gauge bosons~\cite{Blumlein:1994qd,Blumlein:1996qp}. However, we ignore direct electroweak $\chi$-$V$ couplings in our analysis.} We take two benchmark cases with $\kp=0$ and $\kp=1$ in our analysis.


\section{LHC Phenomenology \& Search Strategy}
\label{sec:pheno}
\noindent
We keep our computational setup the same as before~\cite{Chandak:2019iwj}.  We use  {\sc FeynRules}~\cite{Alloul:2013bka} to 
create the UFO~\cite{Degrande:2011ua}
model files for the Lagrangians in Eqs.~\eqref{eq:simplelag1}--\eqref{eq:simplelag5}. Both the signal and background events 
are generated in {\sc MadGraph5}~\cite{Alwall:2014hca} at the leading order (LO). We include 
higher-order corrections to the background processes with QCD $K$ factors wherever available. 
For VLQs, higher-order $K$ factors for signal processes are  not yet known.
We use NNPDF2.3LO~\cite{Ball:2012cx} parton distribution 
functions  with default dynamical renormalization and factorization scales to generate events in
{\sc MadGraph5} and then pass them through {\sc Pythia6}~\cite{Sjostrand:2006za} for showering and 
hadronization. 
Detector effects are simulated using 
{\sc Delphes3}~\cite{deFavereau:2013fsa} with the default CMS card.
Fatj ets are reconstructed from {\sc Delphes}  
tower objects using the Cambridge-Achen~\cite{Dokshitzer:1997in} clustering algorithm (with $R = 1.5$) in 
{\sc FastJet}~\cite{Cacciari:2011ma}. We reconstruct 
hadronic tops from fat jets with {\sc HEPTopTagger}~\cite{Plehn:2010st} with default parameters except for the top-mass window which we relax a little to $80$ GeV from the default $50$ GeV to keep more signal events.

\subsection{Production at the LHC}

\begin{figure*}[]
\captionsetup[subfigure]{labelformat=empty}
\subfloat[\quad\quad\quad(a)]{\includegraphics[width=\columnwidth]{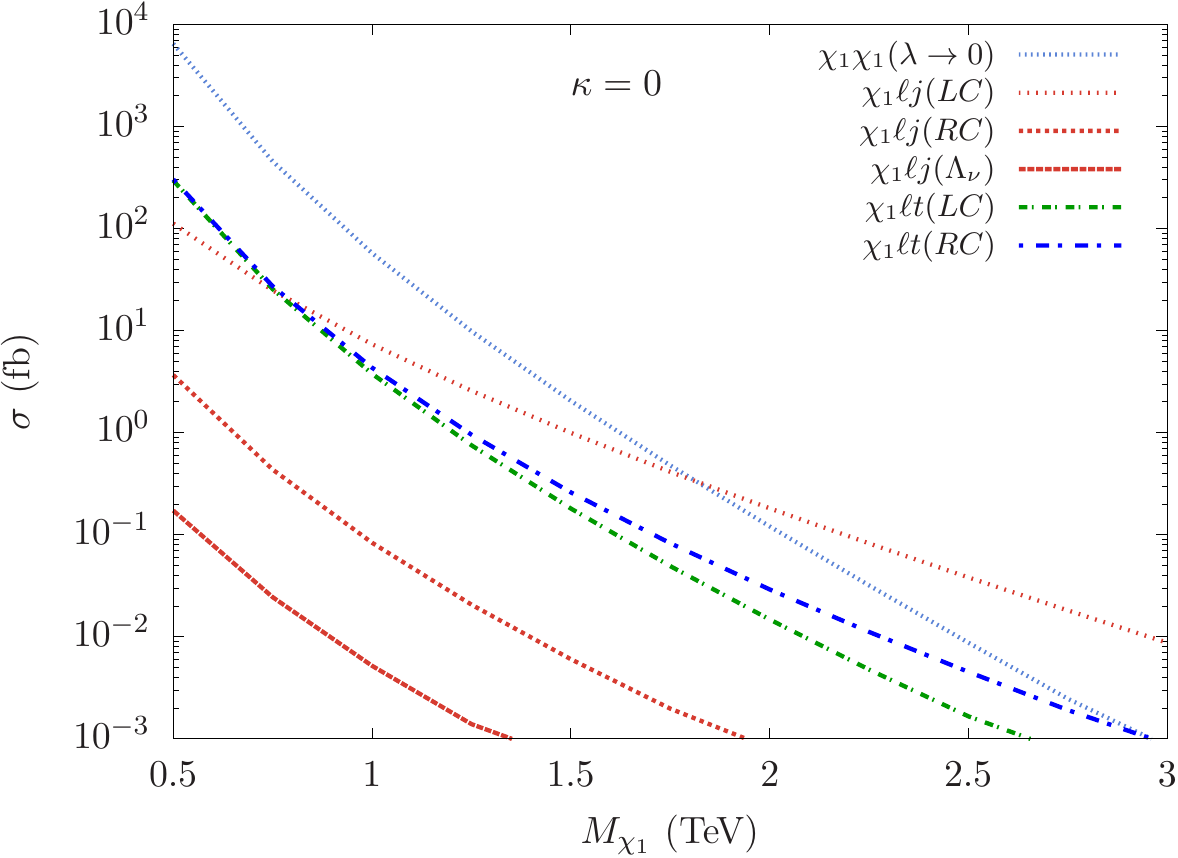}\label{fig:xsec01}}\quad
\subfloat[\quad\quad\quad(b)]{\includegraphics[width=\columnwidth]{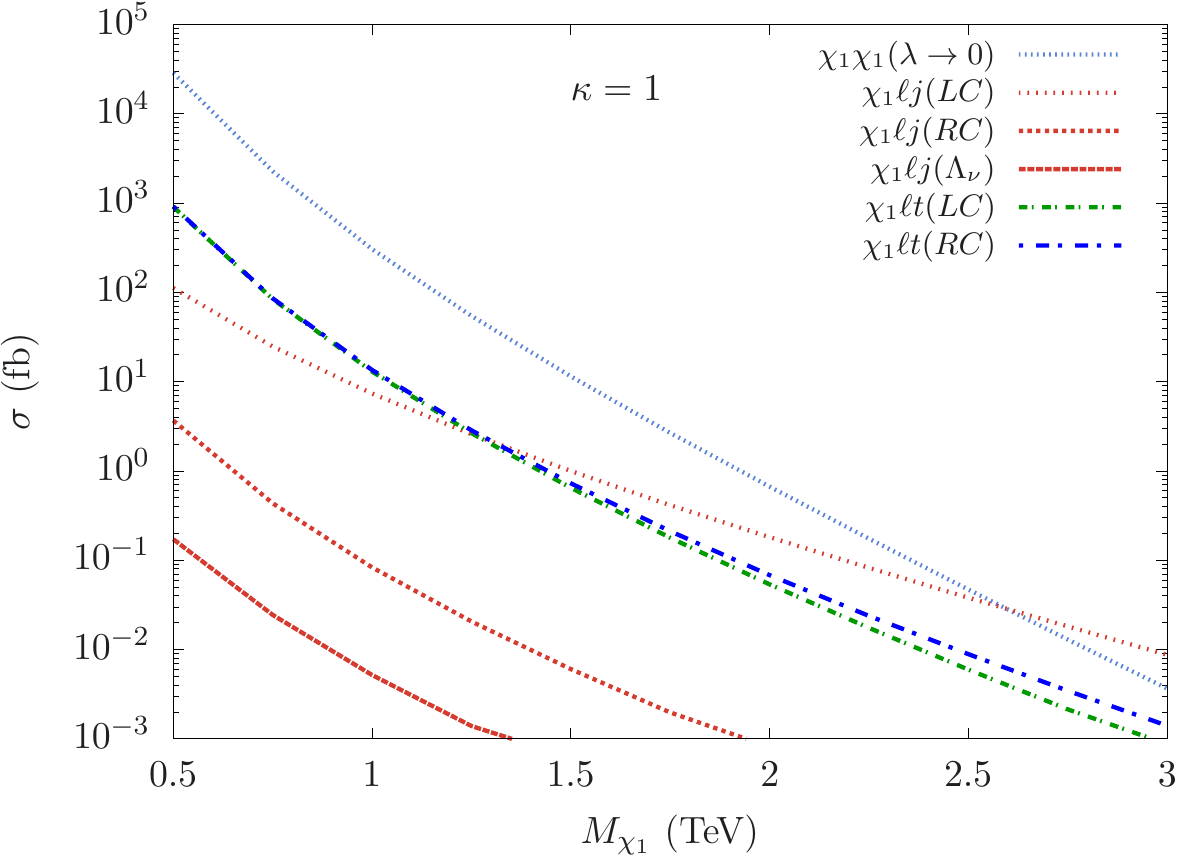}\label{fig:xsec11}}\\
\subfloat[\quad\quad\quad(c)]{\includegraphics[width=\columnwidth]{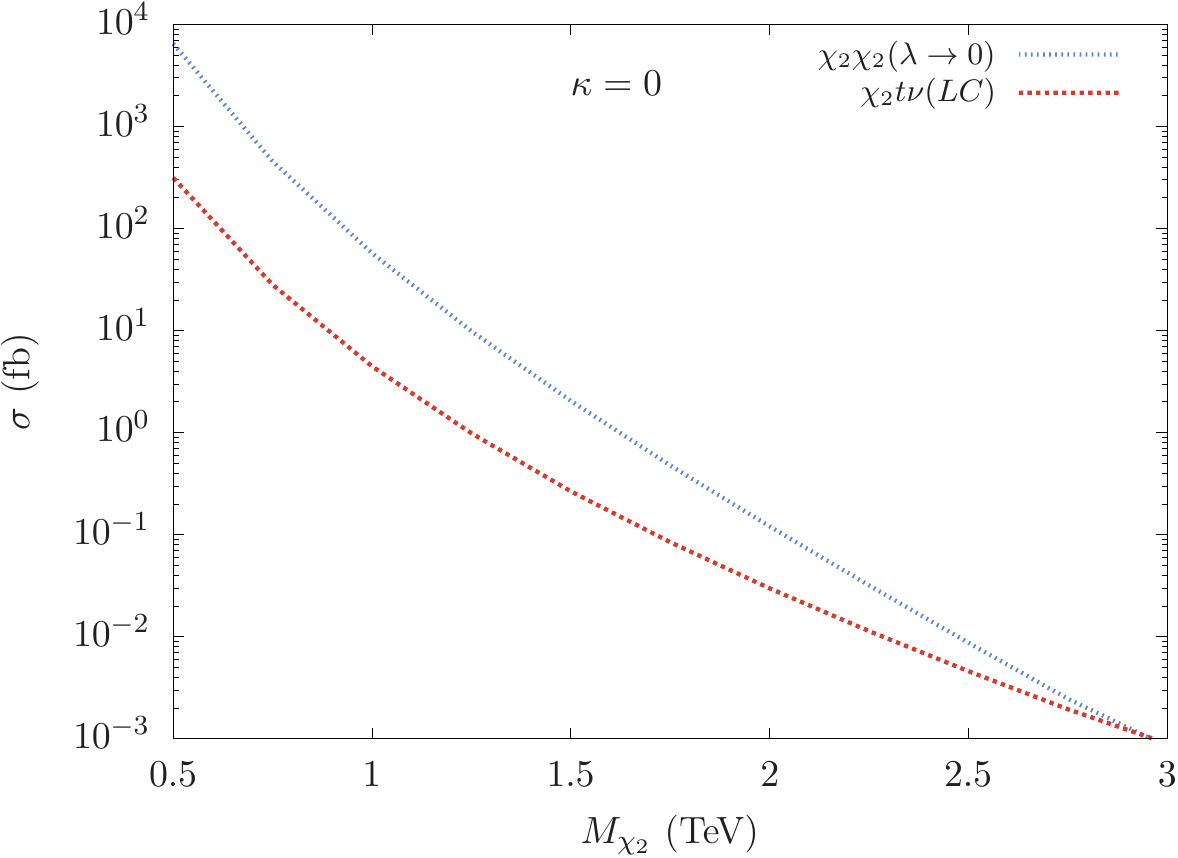}\label{fig:xsec02}}\quad
\subfloat[\quad\quad\quad(d)]{\includegraphics[width=\columnwidth]{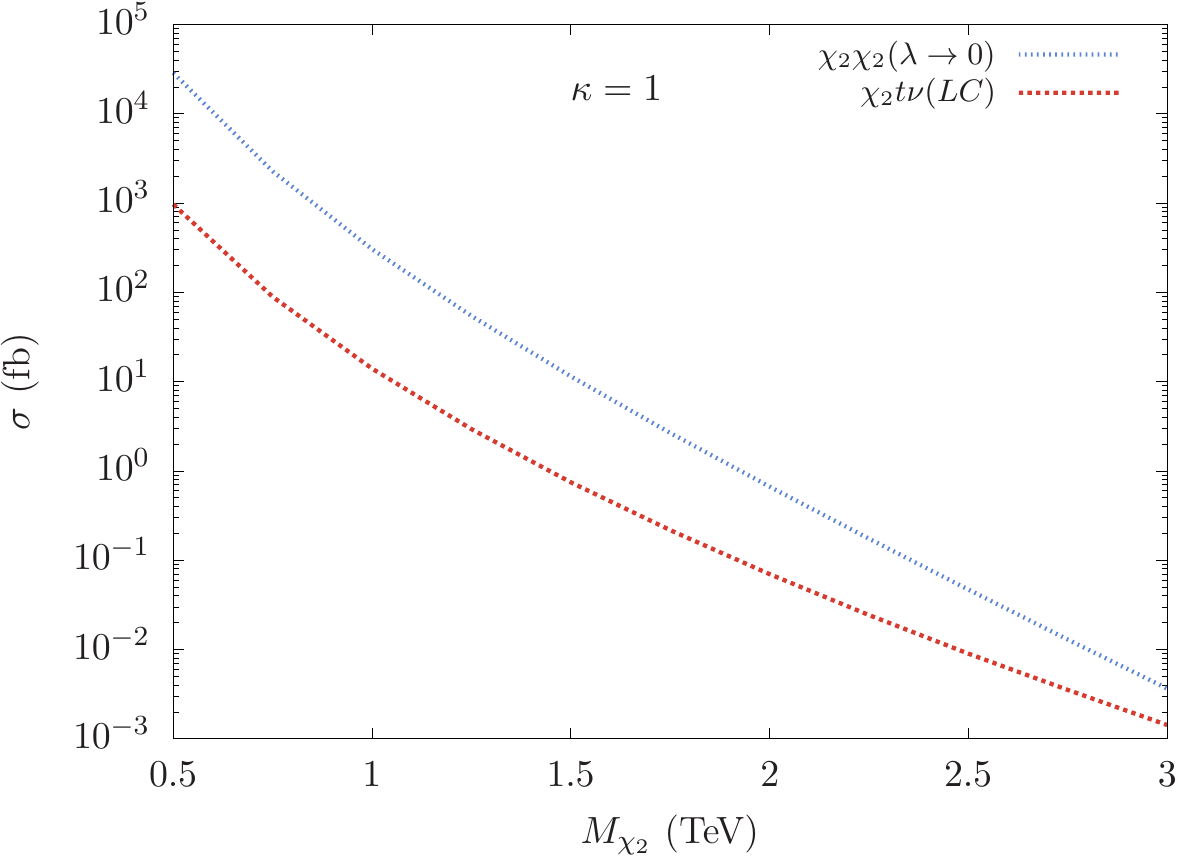}\label{fig:xsec12}}\\
\subfloat[\quad\quad\quad(e)]{\includegraphics[width=\columnwidth]{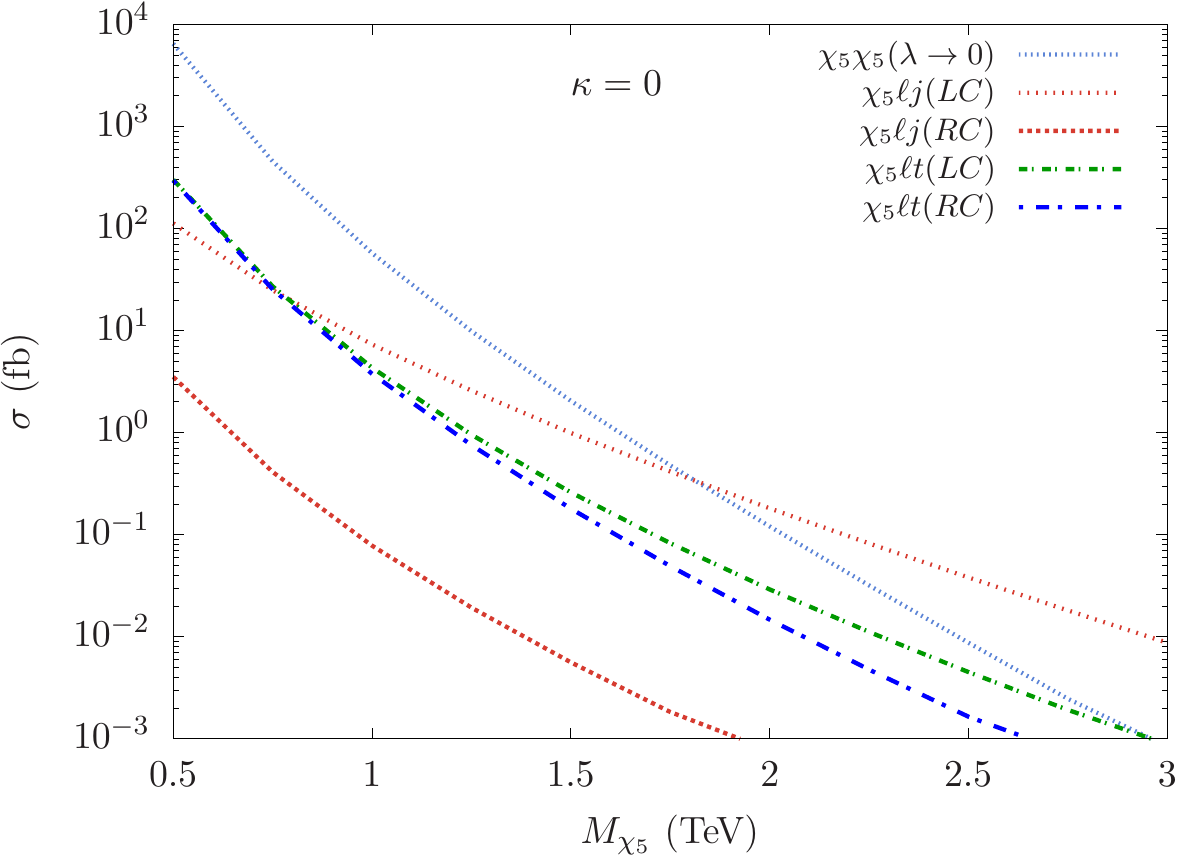}\label{fig:xsec05}}\quad
\subfloat[\quad\quad\quad(f)]{\includegraphics[width=\columnwidth]{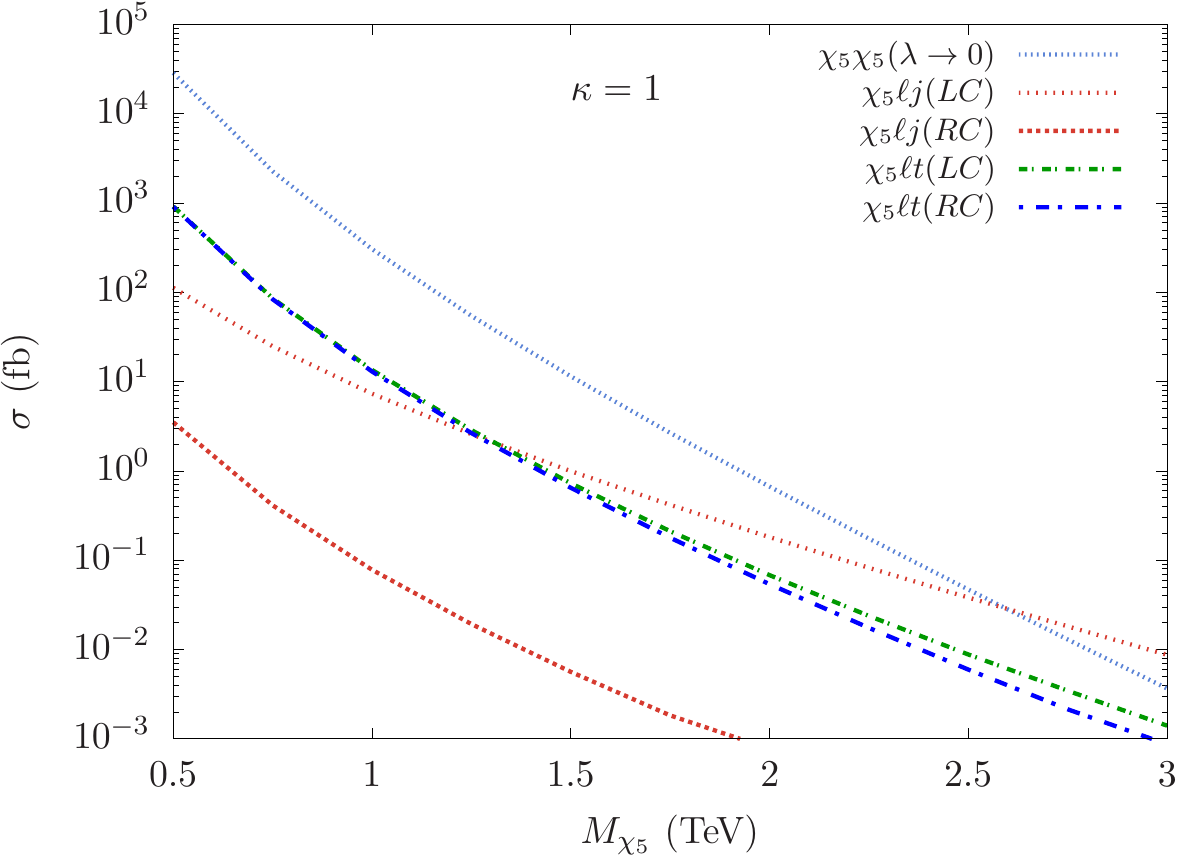}\label{fig:xsec15}}\\
\caption{The parton-level cross sections of different production channels of $\chi_1$ [(a) and (b)], $\chi_2$ [(c) and (d)], and $\chi_5$ 
[(e) and (f)] at the $14$ TeV LHC as functions of $M_{\chi_n}$. The single production cross sections are computed for a benchmark coupling 
$\lm=1$ (see Table~\ref{tab:benchmark}).  Here, $\ell$ stands for either an electron or a muon and the $j$ in the single production processes includes all the light jets as well 
as $b$ jets. Their cross sections are generated with a cut on the transverse momentum of the jet, $p^j_{\rm T} > 20$ GeV.}
\label{fig:xsec}
\end{figure*}

\noindent
The vLQs would be produced resonantly at the LHC through the pair and the single
production channels.  
The dominant pair production diagrams are free of the new couplings and depend only on the universal strong coupling (there are diagrams with $t$-channel lepton exchange that involve new couplings [see Fig.~\ref{fig:feyndiag11}], but their contribution to the total pair production cross section is small~\cite{Mandal:2015vfa}); hence, the process is mostly model independent up to a choice of $\kp$. The pair production would lead to the following final states:
\begin{eqnarray}
pp\ \to\left\{\begin{array}{ccl}
 \chi_1\chi_1&\to& (t\ell)(t\ell)\ /\ (t\ell)(b\n)\ /\ (b\n)(b\n)\\
 \chi_2\chi_2&\to& (t\n)(t\n)\ /\ (t\n)(b\ell)\ /\ (b\ell)(b\ell)\ \\
 \chi_5\chi_5&\to& (t\ell)(t\ell)
\end{array}\right\}.~\label{eq:pair_all}
\end{eqnarray}
Here, as we did for the sLQs~\cite{Chandak:2019iwj}, we ignore those channels with no top quark and consider only symmetric channels i.e., both of the vLQs decay to the same final state. 
Constraining ourselves to such channels will restrict the possible SM backgrounds and make our signal easier to detect. It is generally believed that the symmetric modes
have good discovery prospects~\cite{Diaz:2017lit}.\footnote{The asymmetric modes (where the two LQs decay differently) have not been used for LQ searches so far. For some LQ models, asymmetric channels could provide a better reach than symmetric channels and, therefore, require a separate dedicated analysis~\cite{BMM:2020}.} With these considerations, we are now left with
only the $(t\ell)(t\ell)$ (for $\chi_1$ or $\chi_5$) and $(t\n)(t\n)$ (for $\chi_2$) channels.

With similar consideration for the single production processes, where a LQ is produced in association with a lepton and either a jet or a top quark, the possible final states are given as
\begin{eqnarray}
pp&\ \to&\left\{\begin{array}{ccl}
\chi_1 t \ell &\to & (t\ell)t\ell \\
\chi_1\ell j &\to &  (t\ell)\ell j
\end{array}\right\},\label{eq:single1}\\
pp&\ \to&\left\{\begin{array}{ccl}
\chi_2 t \n &\to & (t\n)t\n \\
\chi_2\n j &\to &  (t\nu)\n j
\end{array}\right\},\label{eq:single2}\\
pp&\ \to&\left\{\begin{array}{ccl}
\chi_5 t \ell &\to & (t\ell)t\ell \\
\chi_5\ell j &\to &  (t\ell)\ell j
\end{array}\right\}.\label{eq:single5}
\end{eqnarray}
In Fig.~\ref{fig:feyndiag} we show some representative Feynman diagrams of the pair and single productions of vLQs.

In Fig.~\ref{fig:xsec} we show the parton-level cross sections of different production processes of $\chi_1$ 
[Figs.~\ref{fig:xsec01}--\ref{fig:xsec11}],  $\chi_2$ [Figs.~\ref{fig:xsec02},--\ref{fig:xsec12}] and $\chi_5$ [Figs.~\ref{fig:xsec05},--\ref{fig:xsec15}] as functions of their masses. The single production cross sections scale as $\lm^2$. Here they are computed for a different benchmark scenarios with reference value $\lm=1$.
We see that in the LC scenario with $\kp=0$, the single production cross section $\sg(pp\to\chi_1\ell j)$ overtakes the pair production cross section at about $1.8$ TeV, while  $\sg(pp\to\chi_1 t\ell)$ always remains smaller. For $\kp=1$, the pair production cross section increases, moving the crossover point with $\sg(pp\to\chi_1\ell j)$ to about $2.6$ TeV. Interestingly, we find that $\sg(pp\to\chi_1 t\ell )$ depends on the choice of $\kp$ despite being a single production process as it contains the $\kp$-dependent $\chi_1\chi_1 g$ vertex. 
In the RC scenario, $\sg(pp\to\chi_1\ell j)$ is reduced by almost $2$ orders of magnitude compared to that in the LC scenario. This happens because in the RC scenario, a $\chi_1$ 
couples to a right-handed top that comes from another left-handed top generated in the charged-current interaction through a chirality flip.
If the $\Lm_\nu$ coupling alone is turned on, the cross section for the $pp\to\chi_1\ell j$ process is negligible [see Figs.~\ref{fig:xsec01}--\ref{fig:xsec11}]. (Note, however, that a nonzero $\Lm_\nu$ can still affect the BRs. For example, we can consider RLCSS and RLCOS scenarios where the BRs for the $\chi_1\to t\ell$ and $\chi_1\to b\nu$ modes are $50$\% each.)
Now, because of the small contribution from the $\Lm_\n$-dependent diagrams and the fact that there is no interference in both the RLCSS and RLCOS scenarios, the $pp\to\chi_1\ell j$ process would have the same cross section as in the RC scenario.
For $\chi_2$, pair production $pp\to\chi_2\chi_2$ always dominates over single production $pp\to \chi_2 t\nu$ up to a mass of $3$ TeV with $\lm=1$ coupling for both $\kp=0$ and
$\kp=1$. In this case, we obtain a $tt$ plus large $\slashed E_T$ signature which was analyzed in Ref.~\cite{Vignaroli:2018lpq}. 
The $\chi_5$ vLQ is similar to the $\chi_1$ and yields similar signatures at the colliders.

The distinctive feature of our signal is the presence of boosted top quarks and high-$p_T$ charged leptons. In symmetric modes, we have at least one top quark in the final state for 
single production while the pair production gives rise to two top quarks. In both cases, we have two high-$p_{\rm T}$ charged leptons. Therefore, as already indicated in the \hyperref[sec:intro]{Introduction}, we
combine events from both pair and single productions by demanding at least one top jet (a hadronically decaying top quark forming a fat jet) and exactly two high-$p_{\rm T}$ same-flavor-opposite-sign (SFOS) leptons in the final state to enhance the signal sensitivity.
Note that the same final state can arise from both pair and single productions. For example, the $t\ell t\ell$ state can come from both $pp\to \chi_{1,5}\chi_{1,5}$ and 
$pp\to \chi_{1,5}t\ell$ processes (see Fig.~\ref{fig:feyndiag}). This can lead to double counting the contribution of some diagrams while generating signal events. 
One can avoid this by ensuring that both $\chi$ and $\chi^\dag$ are not 
on-shell simultaneously in any single production event~\cite{Mandal:2015vfa}.

\begin{table}[t!]
\begin{center}
\begin{tabular}{|c|c|c|c|}
\hline
\multicolumn{2}{|c|}{Background } & $\sg$ & QCD\\ 
\multicolumn{2}{|c|}{processes}&(pb)&Order\\\hline\hline
$V +$ jets  & $Z +$ jets  &  $6.33 \times 10^4$& NNLO \\ \cline{2-4} 
 \cite{Catani:2009sm,Balossini:2009sa}                 & $W +$ jets  & $1.95 \times 10^5$& NLO \\ \hline
$VV +$ jets  & $WW +$ jets  & 124.31& NLO\\ \cline{2-4} 
    \cite{Campbell:2011bn}               & $WZ +$ jets  & 51.82 & NLO\\ \cline{2-4} 
                   & $ZZ +$ jets  &  17.72 & NLO\\ \cline{1-4}
Single $t$  & $tW$  &  83.1 & N$^2$LO \\ \cline{2-4} 
\cite{Kidonakis:2015nna}                   & $tb$  & 248.0 & N$^2$LO\\ \cline{2-4} 
                   & $tj$  &  12.35 & N$^2$LO\\  \cline{1-4}
$tt$~\cite{Muselli:2015kba}  & $tt +$ jets  & 988.57 & N$^3$LO\\ \cline{1-4}
\multirow{2}{*}{$ttV$~\cite{Kulesza:2018tqz}} & $ttZ$  &  1.045 &NLO+NNLL \\ \cline{2-4} 
                   & $ttW$  & 0.653& NLO+NNLL \\ \hline
\end{tabular}
\caption{Total cross sections without any cut for SM background processes considered in our analysis. The higher-order  QCD cross sections are taken from the literature and are shown in the last column. We use these cross sections to compute the $K$ factors which multiply the LO cross sections to include higher-order effects.}
\label{tab:Backgrounds}
\end{center}
\end{table}

\begin{figure*}[]
\captionsetup[subfigure]{labelformat=empty}
\subfloat[\quad\quad\quad(a)]{\includegraphics[width=\columnwidth]{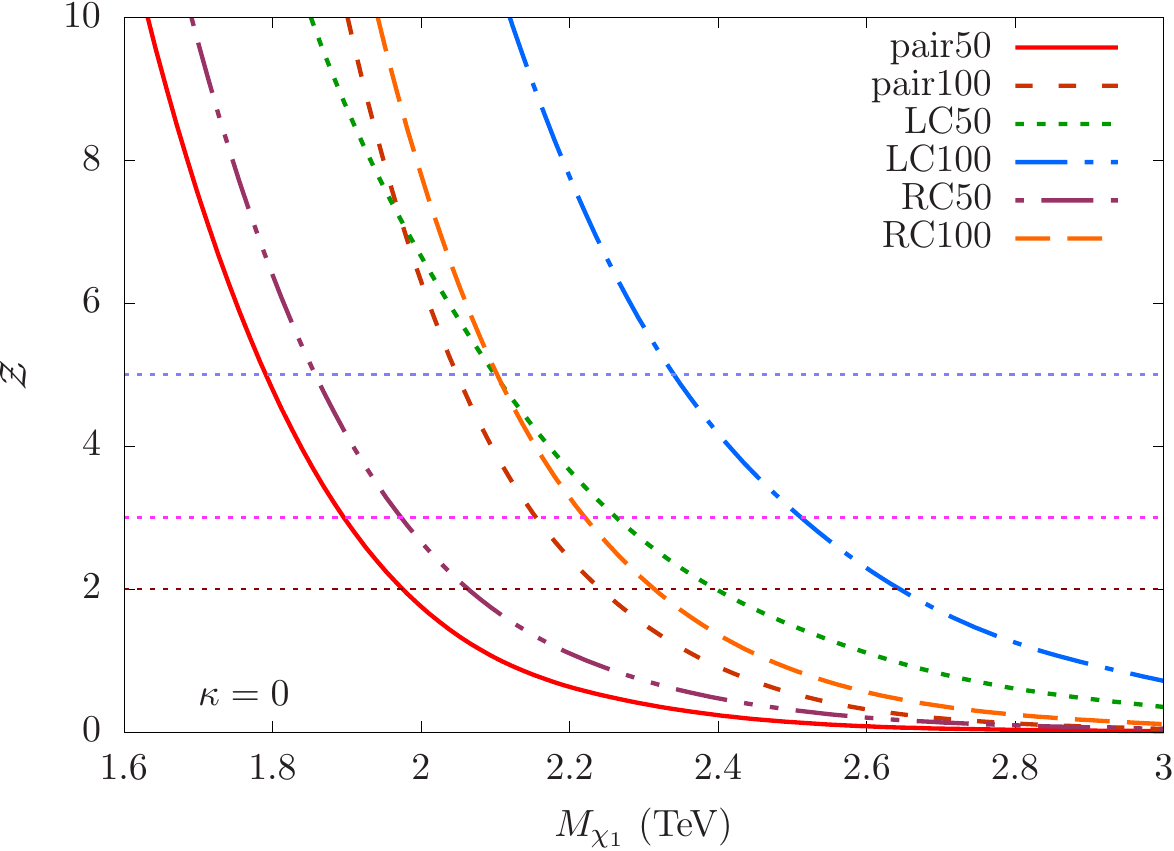}\label{fig:xset01}}\quad
\subfloat[\quad\quad\quad(b)]{\includegraphics[width=\columnwidth]{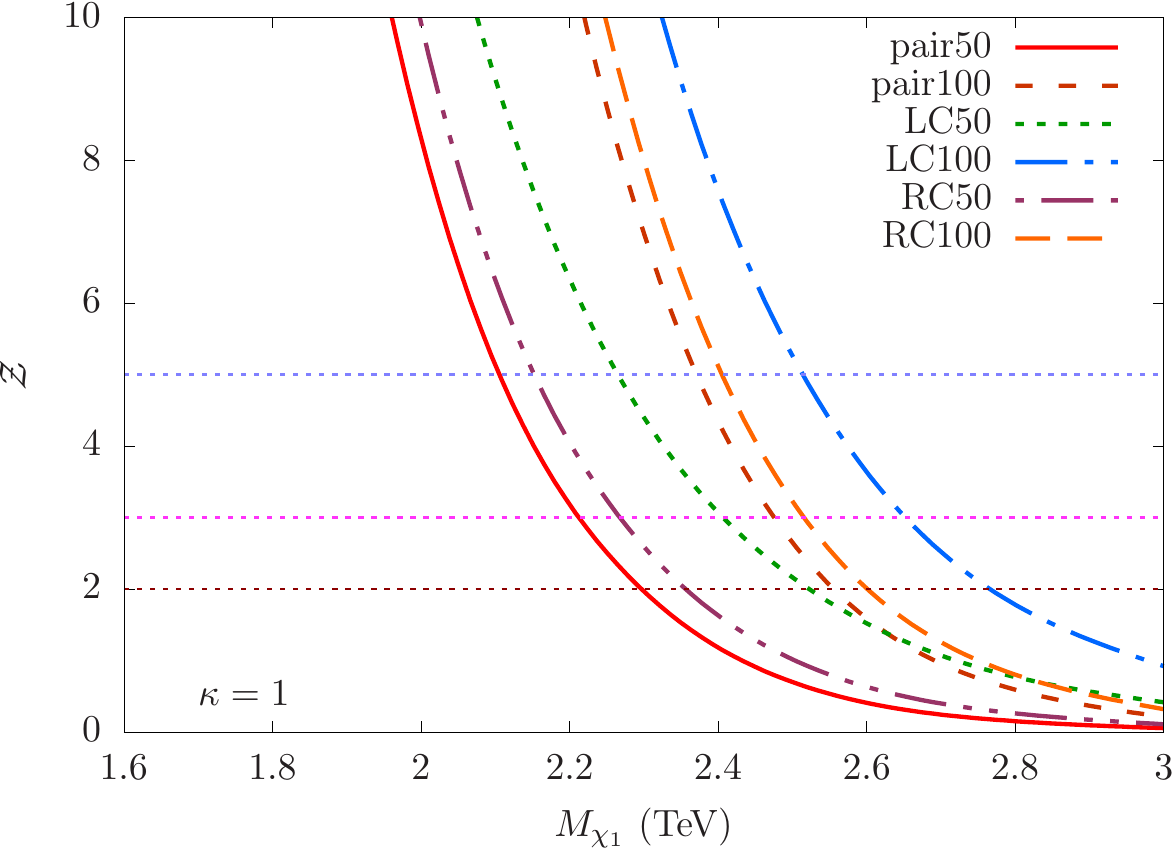}\label{fig:xset11}}\\
\subfloat[\quad\quad\quad(c)]{\includegraphics[width=\columnwidth]{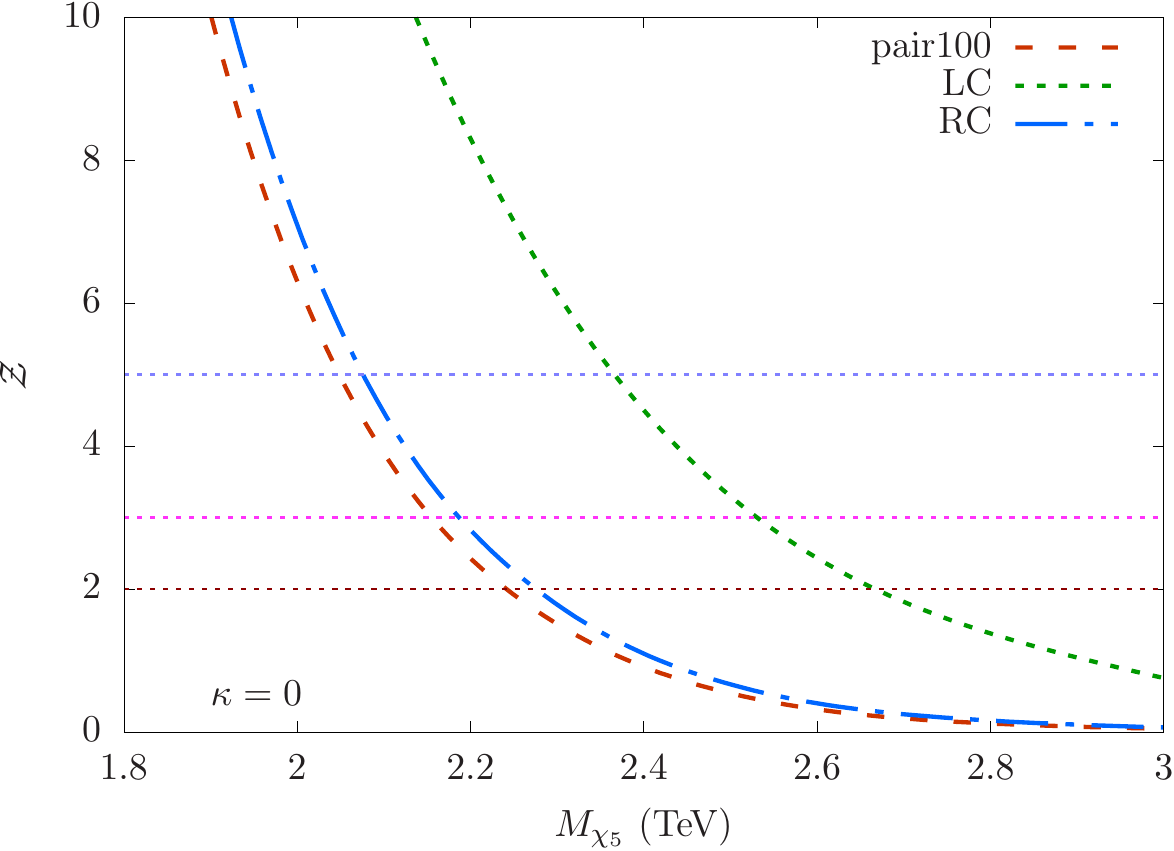}\label{fig:xset02}}\quad
\subfloat[\quad\quad\quad(d)]{\includegraphics[width=\columnwidth]{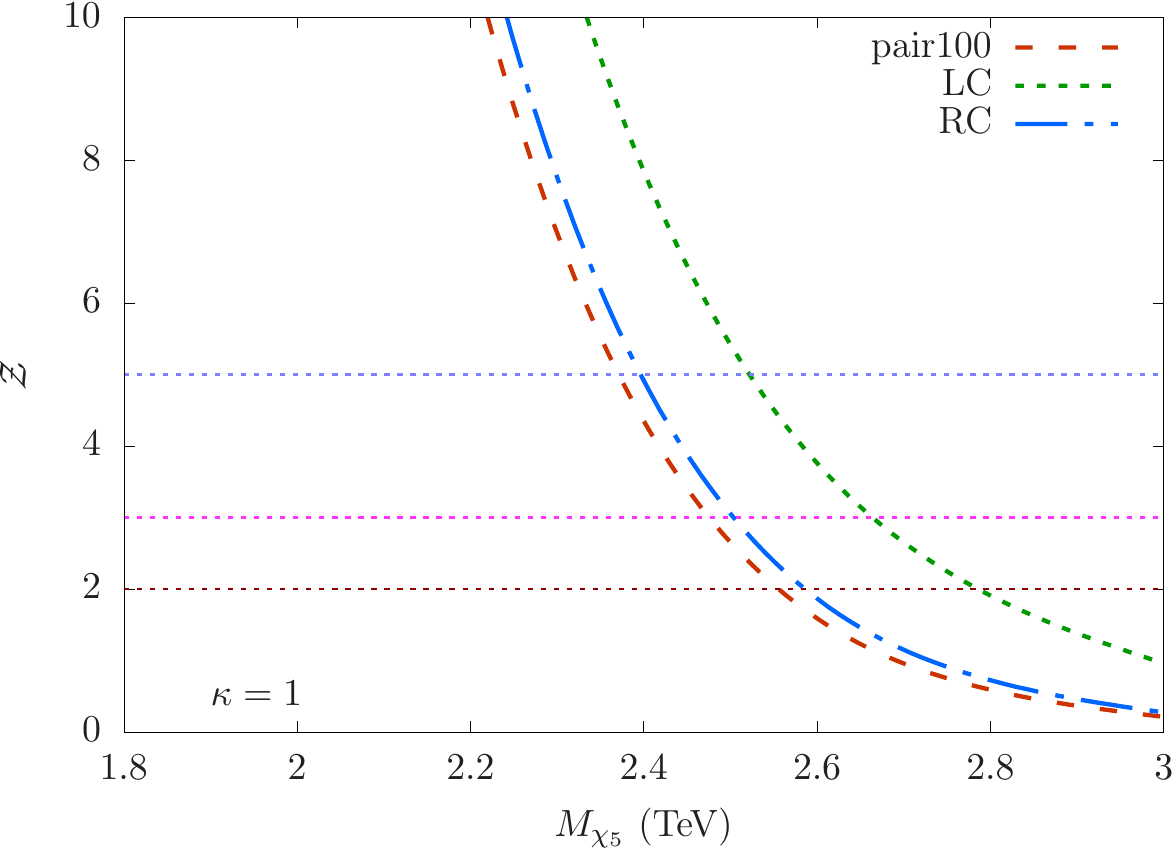}\label{fig:xset12}}\\
\caption{Expected significance $\mc Z$ in units of standard deviation $\sg$ for observing the $\chi_{1}$ (a)[$\kappa=0$],(b)[$\kappa=1$] and $\chi_{5}$(c)[$\kappa=0$],(d)[$\kappa=1$] signals over the SM backgrounds. They are plotted as functions of their masses for $3$ ab$^{-1}$ of integrated luminosity at the 14 TeV HL-LHC for different coupling scenarios in the electron mode. We use the combined pair and single productions for the signals in the LC and RC scenarios. We also show the pair production significance for 50\% and 100\% BRs in the $\chi\rightarrow t\ell$ decay mode. We have considered $\lambda=1$ when computing the signals.}
\label{fig:xset}
\end{figure*}

\begin{table*}[]
\begin{tabular}{|p{0.5cm} || c c c c | c c | c c | c || c c c c | c c | c c | c |}
\hline
 \multirow{3}{*}{~\rotatebox[origin=c]{90}{Significance $\mc Z$ }} & \multicolumn{18}{c|}{Limit on $M_\chi$ (TeV)}\\ 
	& \multicolumn{8}{c}{$\kappa=0$}	& \multicolumn{10}{c|}{$\kappa=1$} \\ \cline{2-19} 
                         & \multicolumn{6}{c|}{$\chi_1$} & \multicolumn{3}{c||}{$\chi_5$} & \multicolumn{6}{c|}{$\chi_1$} & \multicolumn{3}{c|}{$\chi_5$} \\ 
\cline{2-19}
&\multicolumn{4}{c|}{Combined}&\multicolumn{2}{c|}{Pair}&\multicolumn{2}{c|}{Combined}&Pair&\multicolumn{4}{c|}{Combined}&\multicolumn{2}{c|}{Pair}&\multicolumn{2}{c|}{Combined}&Pair\\
\cline{2-19}
& LC50 & LC & RC50 & RC & BR=$0.5$ & BR=$1$ & LC & RC & BR=$1$ & LC50 & LC & RC50 & RC & BR=$0.5$ & BR=$1$ & LC & RC & BR=$1$\\
   \hline\hline
~~5 & 2.10 & 2.34 & 1.85 & 2.10 & 1.79 & 2.05 & 2.36 & 2.07 & 2.04 & 2.26 & 2.51 & 2.14 & 2.40 & 2.10 & 2.36 & 2.52 & 2.39 & 2.36 \\ \hline
~~3 & 2.25 & 2.51 & 1.97 & 2.22 & 1.89 & 2.15 & 2.52 & 2.18 & 2.15 & 2.40 & 2.65 & 2.26 & 2.51 & 2.21 & 2.47 & 2.66 & 2.50 & 2.47 \\ \hline
~~2 & 2.39 & 2.64 & 2.06 & 2.31 & 1.97 & 2.23 & 2.66 & 2.27 & 2.23 & 2.52 & 2.76 & 2.35 & 2.59 & 2.29 & 2.55 & 2.78 & 2.58 & 2.55 \\ \hline
\end{tabular}
\caption{The mass limits corresponding to $5\sg$ (discovery), $3\sg$ and $2\sg$ (exclusion) significances ($\mc Z$) for observing the  (a) $\chi_1$ and (b) $\chi_5$ signals over the SM backgrounds for 3 ab$^{-1}$ integrated luminosity at the $14$ TeV LHC with combined and pair-production-only signals. Here, LC (RS) stands for LC100 (RC100).}
\label{tab:sig}
\end{table*}

\begin{figure*}[!t]
\captionsetup[subfigure]{labelformat=empty}
\subfloat[\quad\quad\quad(a)]{\includegraphics[width=\columnwidth]{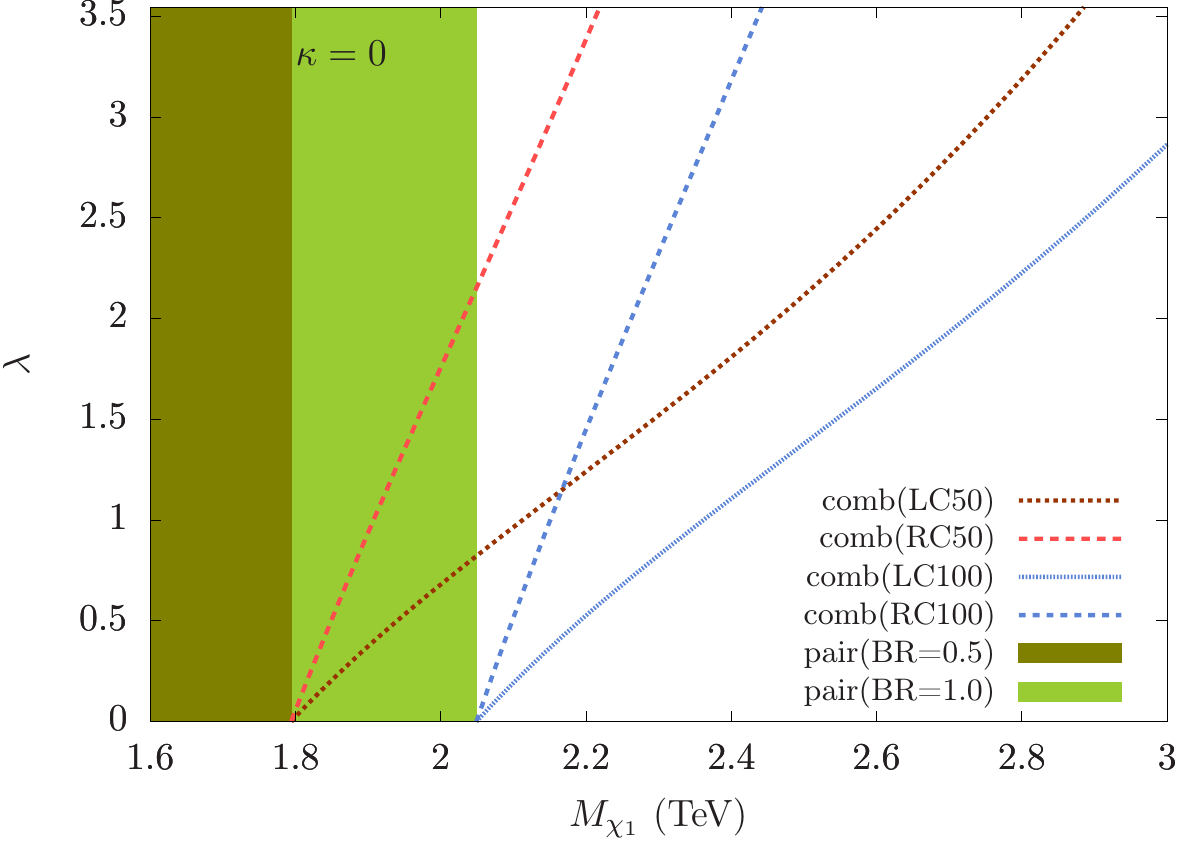}}\label{fig:xlamv1z5r01}\quad
\subfloat[\quad\quad\quad(b)]{\includegraphics[width=\columnwidth]{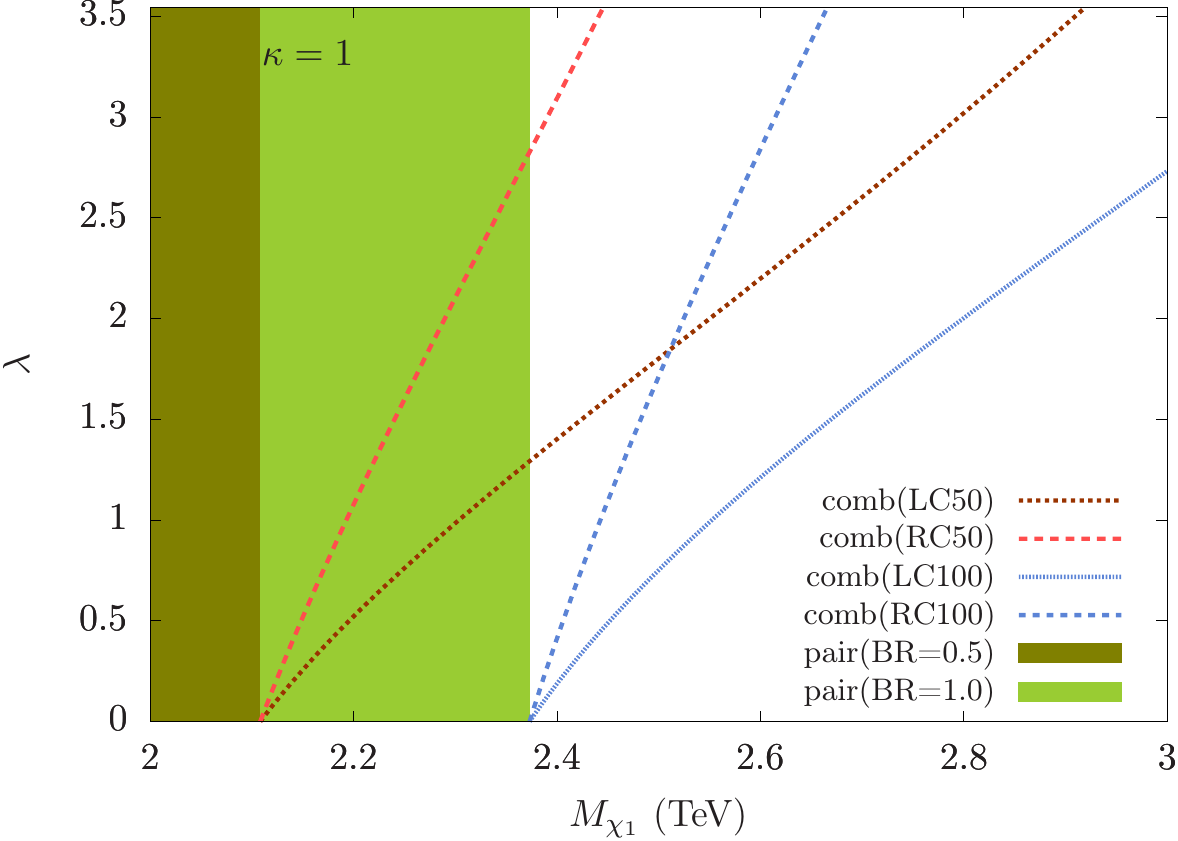}}\label{fig:xlamv1z5r11}\\
\subfloat[\quad\quad\quad(c)]{\includegraphics[width=\columnwidth]{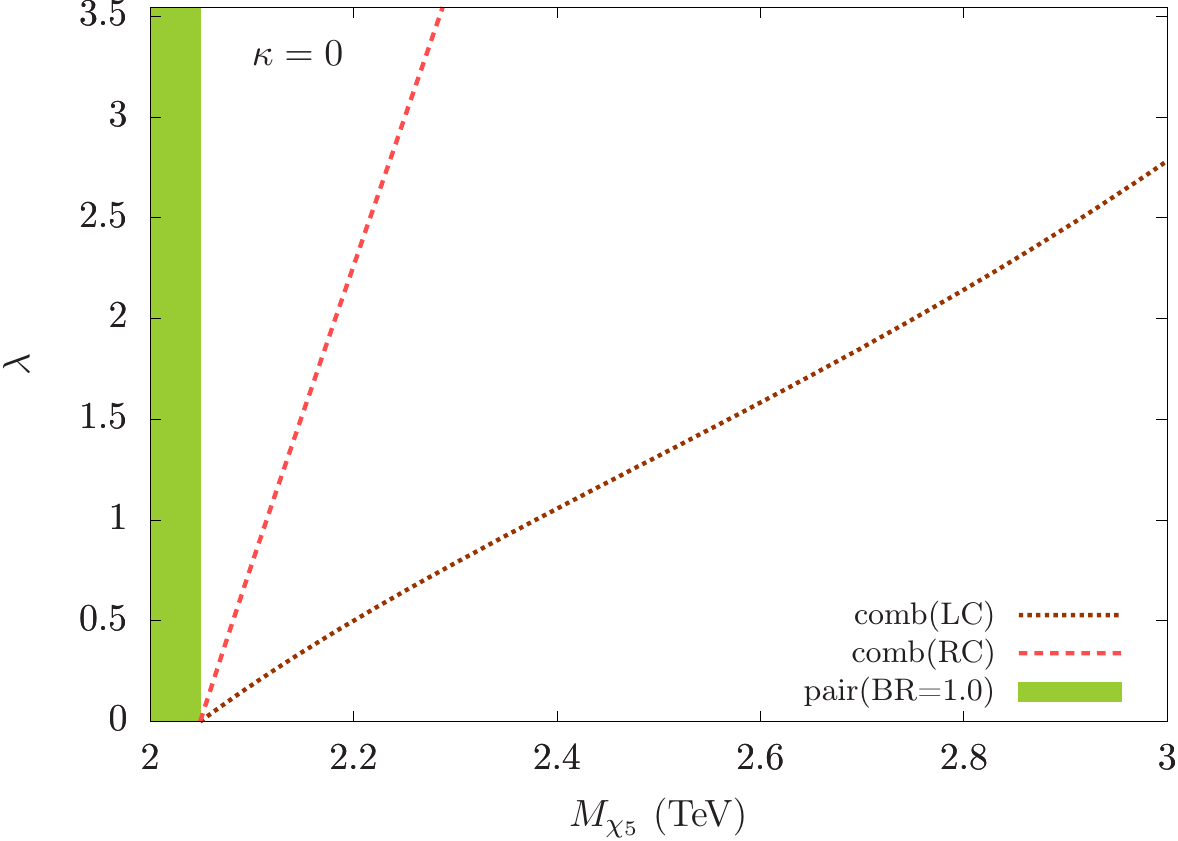}}\label{fig:xlamv1z5r02}\quad
\subfloat[\quad\quad\quad(d)]{\includegraphics[width=\columnwidth]{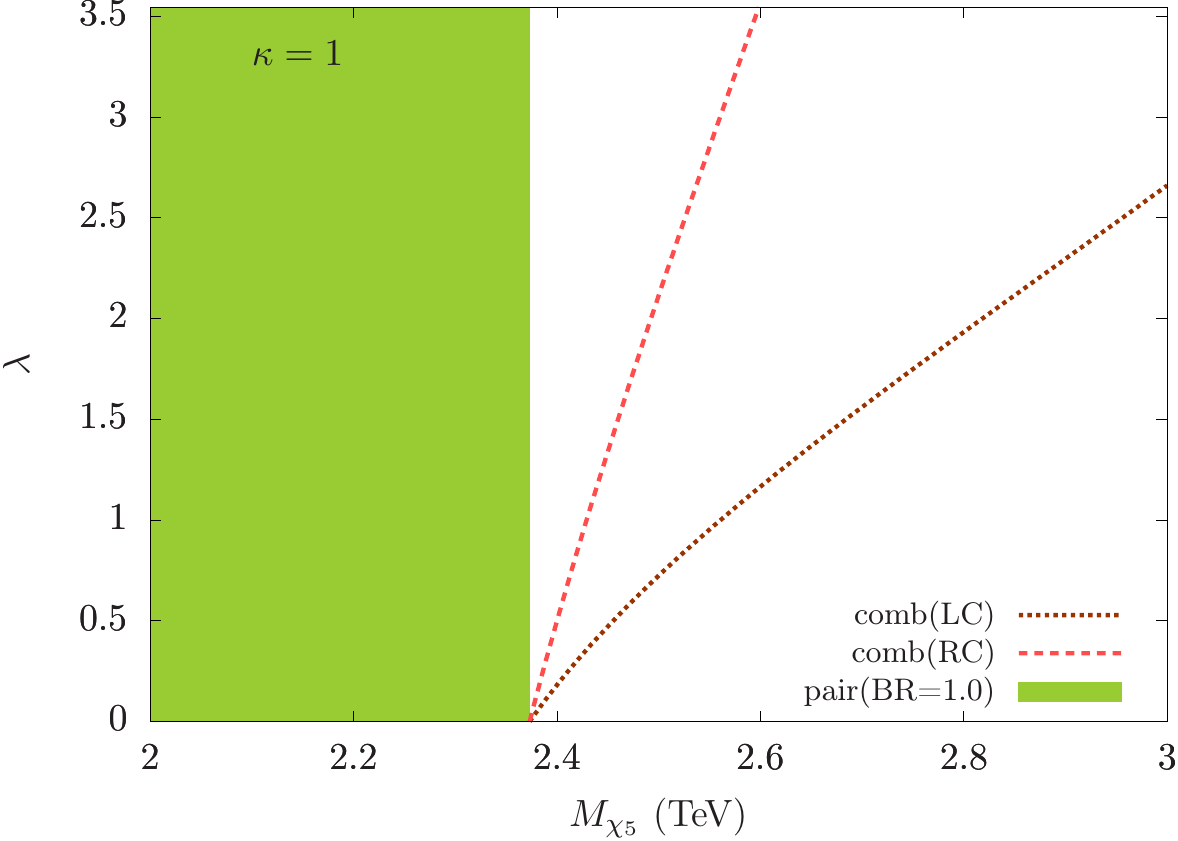}}\label{fig:xlamv1z5r12}\\
\caption{The $5\sg$ discovery reaches in the $\lm$-$M_\chi$ planes for $\chi_1$ with (a) $\kappa=0$ and (b) $\kappa=1$ and for $\chi_5$ with (c) $\kappa=0$ and (d) $\kappa=1$. These plots show the smallest $\lm$ needed to observe  $\chi_1$ and $\chi_5$ signals with $5\sg$ significance for a range of $M_\chi$ with $3$ ab$^{-1}$ of integrated luminosity. The pair-production-only regions for $50\%$ and $100\%$ BRs in the $\chi\to t\ell$ decay mode are shown with shades of green. Since the pair production is insensitive to $\lm$, a small coupling is sufficient to attain $5\sg$ significance within the green regions.}
\label{fig:xlamv1z5r}
\end{figure*}

\begin{figure*}[!t]
\captionsetup[subfigure]{labelformat=empty}
\subfloat[\quad\quad\quad(a)]{\includegraphics[width=\columnwidth]{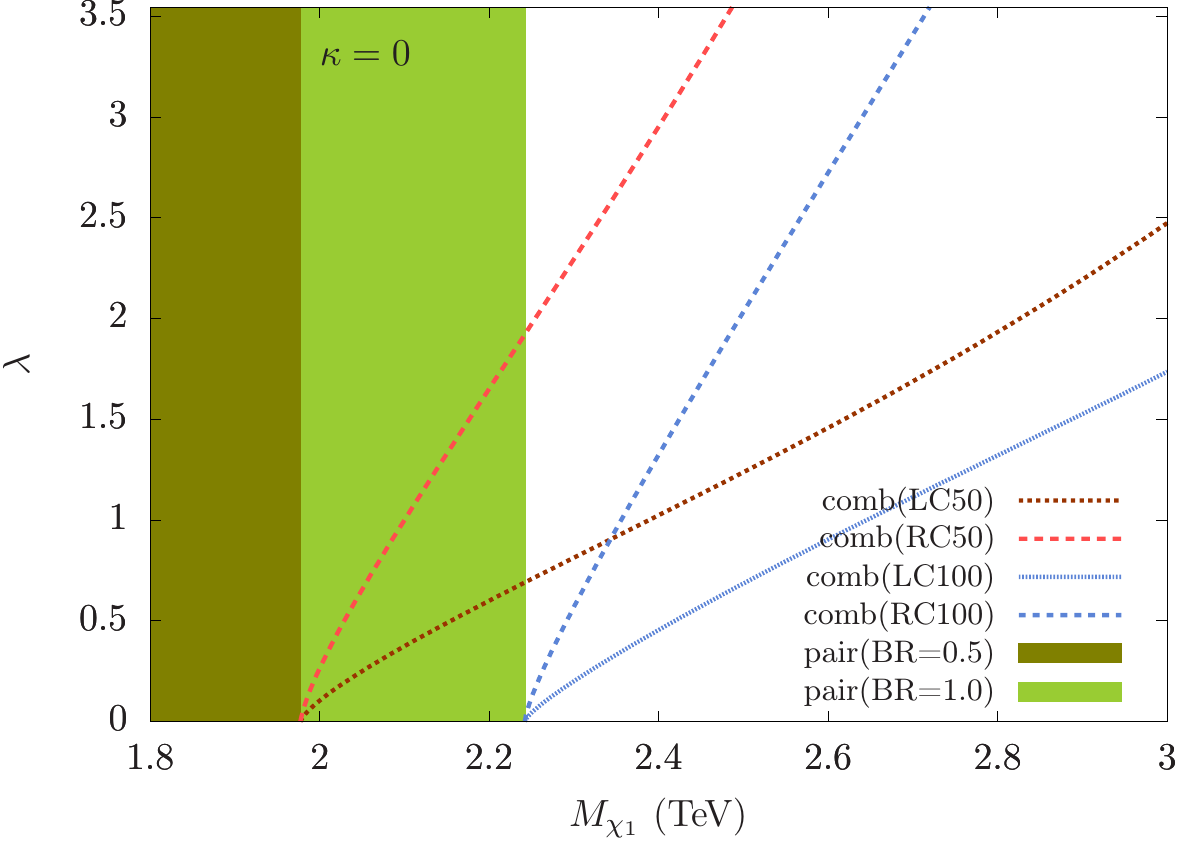}}\label{fig:xlamv1z2s01}\quad
\subfloat[\quad\quad\quad(b)]{\includegraphics[width=\columnwidth]{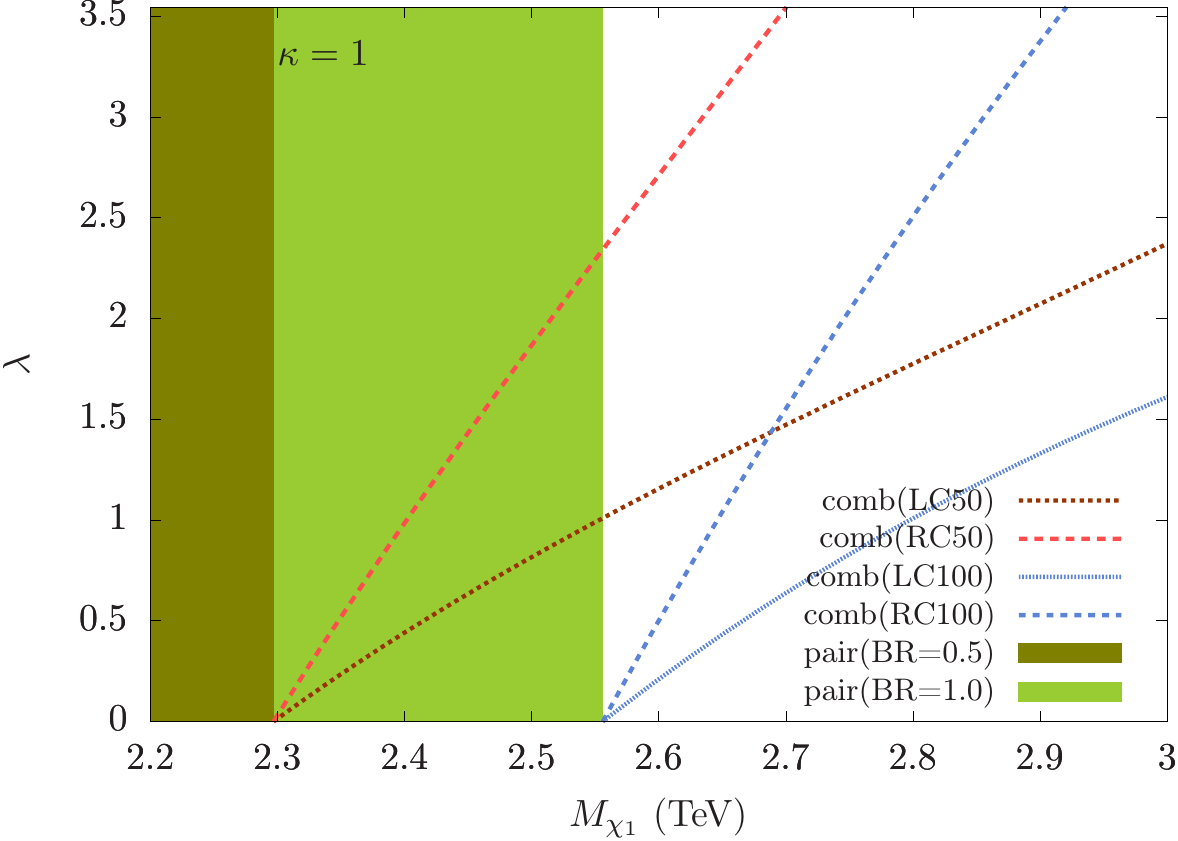}}\label{fig:xlamv1z2s11}\\
\subfloat[\quad\quad\quad(c)]{\includegraphics[width=\columnwidth]{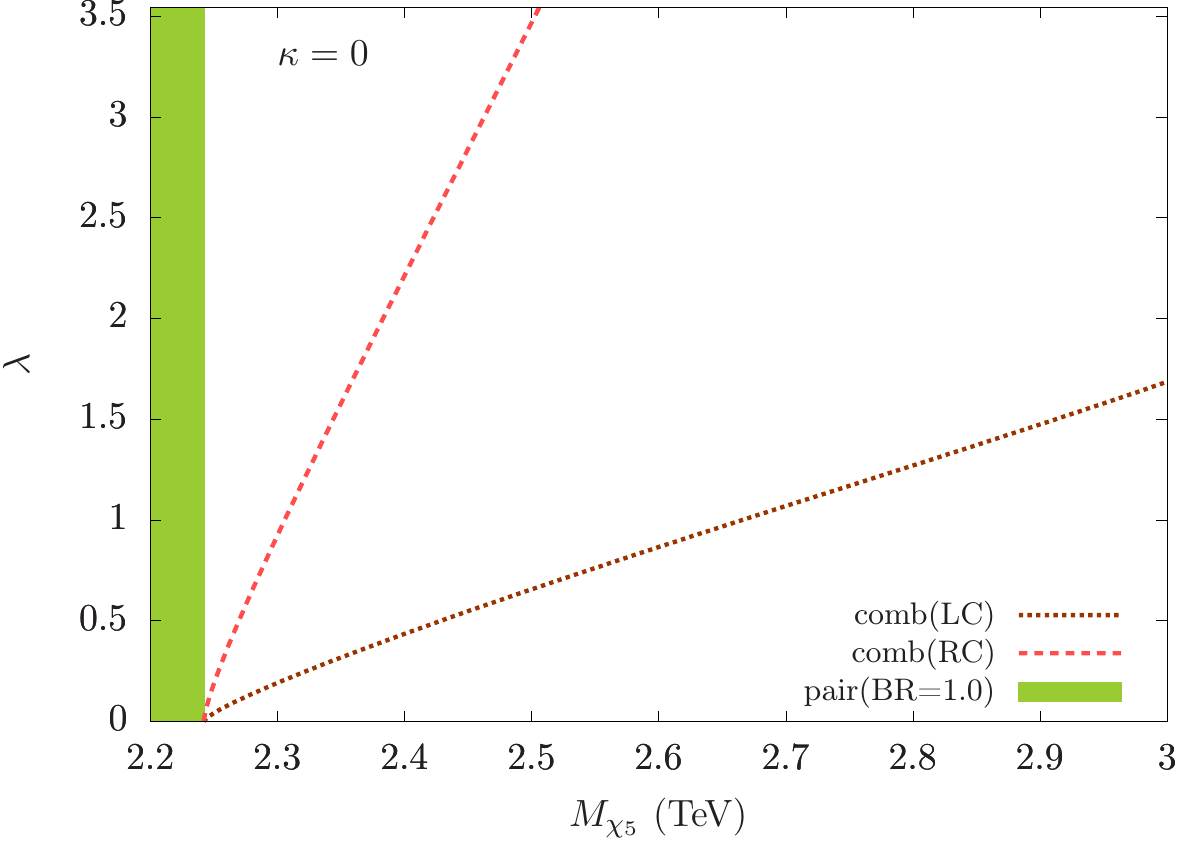}}\label{fig:xlamv1z2s02}\quad
\subfloat[\quad\quad\quad(d)]{\includegraphics[width=\columnwidth]{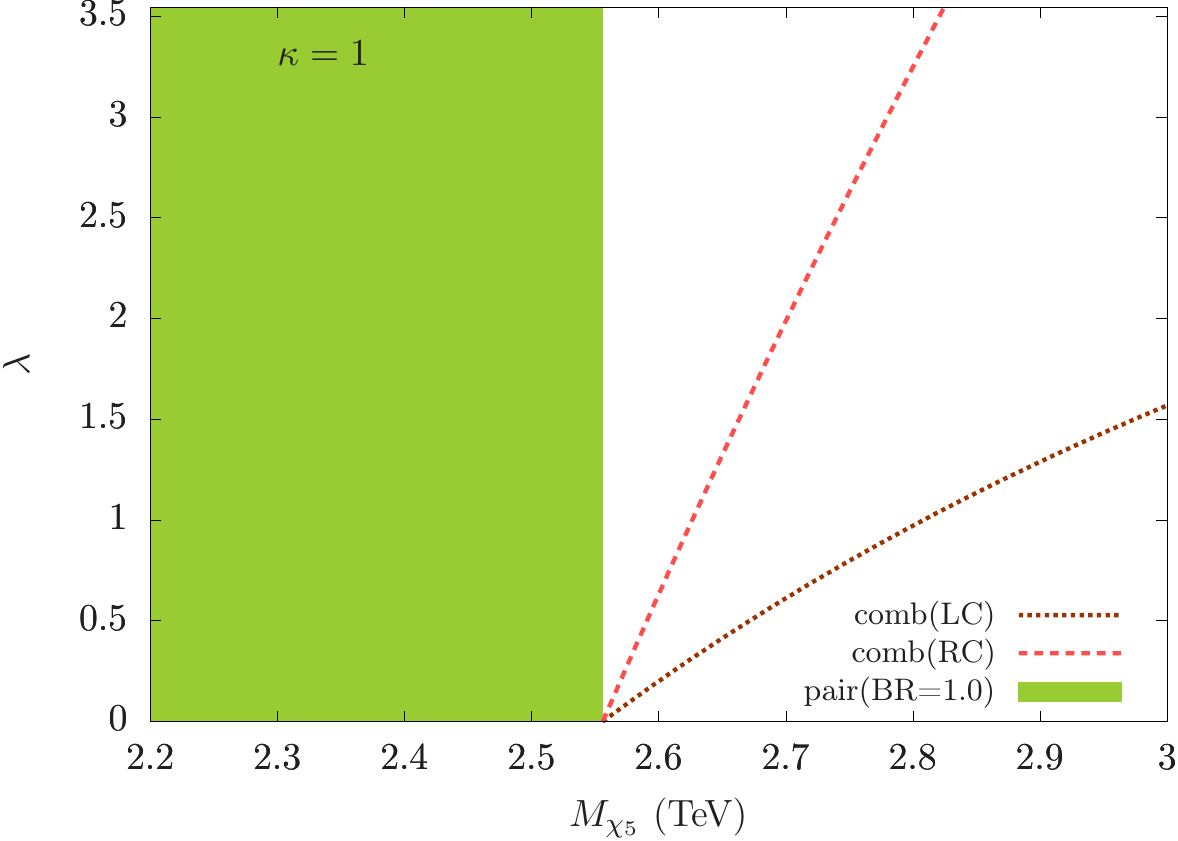}}\label{fig:xlamv1z2s12}\\
\caption{The $2\sg$ exclusion limits in the $\lm$-$M_\chi$ planes for $\chi_1$ with (a) $\kappa=0$ and (b) $\kappa=1$ and for $\chi_5$ with (c) $\kappa=0$ and (d) $\kappa=1$. These plots show the smallest $\lm$ that can be excluded by the HL-LHC  with 3 ab$^{-1}$ of integrated luminosity. The pair-production-only regions for $50\%$ and $100\%$ BRs in the $\chi\to t\ell$ decay mode are shown with green shades.}
\label{fig:xlamv1z2s}
\end{figure*}

\subsection{Backgrounds and selection}

\noindent
Since the topology of the vLQ signal is identical to that of the sLQ signal [i.e., at least one (hadronic) top fatjet and exactly two high-$p_{\rm T}$ SFOS leptons], our background analysis essentially remains the same as before~\cite{Chandak:2019iwj}. Therefore we refer the reader to the earlier paper for a detailed discussion on the possible SM background processes; here, we present the gist of our discussions found there. The dominant SM background processes for our desired signal can arise from processes with two leptons and significant cross sections at the LHC. 
The top-like fat jet can appear from either an actual top quark decaying hadronically or a bunch of QCD/non-QCD jets mimicking its signature. 
We find that $pp\to Z+$jets and $pp\to tt+$jets processes contribute significantly to the background. The single top, diboson, and $ttV$ ($V=W,Z$) production processes are subdominant.
There are SM processes with large cross sections, e.g., $pp\to W+$jets $\to \ell\nu+$jets that can in principle act as backgrounds because of a jet mimicking a lepton. However, we found that
these processes actually contribute negligibly, thanks to a very small misidentification efficiency. 

In Table~\ref{tab:Backgrounds} we list the relevant SM processes and their higher-order cross sections. We consider 
these backgrounds after adjusting with appropriate $K$ factors to include higher-order effects. 
Although the bare cross sections (i.e. without any cut) of some background processes are seemingly huge, we control them by applying strong selection cuts.
These cuts are designed in such a way that they would drastically reduce the background without harming the signal much since our signal possesses specific kinematic
features that are very different from the backgrounds. However, some backgrounds are so big at the beginning (e.g., $Z+$jets) that in order to save computation time and
have better statistics, we apply the following strong cuts at the generation level:
\begin{enumerate}
\item $p_{\rm T}(\ell_1)> 250$ GeV.
\item Invariant mass $M(\ell_1,\ell_2) > 115$ GeV ($Z$-mass veto).
\end{enumerate}
Here $\ell_i$ denotes the $i^{\rm th}$ $p_{\rm T}$-ordered lepton ($e/\m$).
After generating events with the above generation-level cuts, we apply the following final selection criteria sequentially on the signal and background events at the analysis level:

{\renewcommand\theenumi{\bfseries $\mc C_\arabic{enumi}$:}
\renewcommand\labelenumi{\theenumi}

\begin{enumerate}\setcounter{enumi}{0}

\item 
{(a) Minimum one top jet (obtained from {\sc HEPTopTagger}) with $p_{\rm T}(t_h) > 135$ GeV.

(b) Exactly two SFOS leptons with $p_{\rm T}(\ell_1)> 400$ GeV and $p_{\rm T}(\ell_2)> 200$ GeV 
and pseudorapidity $|\eta(\ell)|<2.5$. For $e$, we consider the barrel-end cap cut on $\eta$ between $1.37$ and $1.52$.

(c) Invariant mass of the lepton pair $M(\ell_1,\ell_2) > 120$ GeV ($Z$ veto).

(d) The missing energy $\slashed{E}_T < 200$ GeV.
}
\item The scalar sum of the transverse $p_{T}$ of all visible objects, $S_T>1.2\times {\rm Min}\lt( M_\chi, 1750\rt)$ GeV.

\item Max$\left(M(\ell_1,t)~\mathrm{OR}~M(\ell_2,t)\right) > 0.8\times {\rm Min}\lt(M_{\chi},1750\rt)$ GeV.
\end{enumerate}}

\section{Discovery potential}\label{sec:dispot}

\noindent
We use the following formula to estimate the signal significance $\mc{Z}$:
\begin{align}
\mc{Z} =\sqrt{2\lt(N_S+N_B\rt)\ln\lt(\frac{N_S+N_B}{N_B}\rt)-2N_S}\, ,
\end{align}
where the number of signal and background events surviving the final selection cuts (as listed in the previous section) are denoted by $N_S$ and $N_B$, respectively. 
In Fig.~\ref{fig:xset} we show the expected significance as a function of vLQ masses. As discussed earlier, the choice of $\kp$ affects the pair and some
single productions. In Figs.~\ref{fig:xset01} and \ref{fig:xset11} we present $\mc{Z}$ for $\chi_1$ with $\kappa = 0$ and $\kappa =1$, respectively. Similarly, 
Figs.~\ref{fig:xset02} and \ref{fig:xset12} are for $\chi_5$. These curves are obtained for the $14$ TeV LHC with $3$ ab$^{-1}$ of integrated luminosity.
We have used $\lm=1$  to estimate the significance for the combined signal (i.e. the pair and single production events together). We note the following points:
\begin{itemize}
\item 
The LC100 (RC100) curves for $\chi_1$ and the LC (RC) curves for $\chi_5$ represent the significances in the LC (RC) scenario where the BR of the $\chi_{1}\rightarrow t\ell$ decay is $100$\%. 
\item 
For $\chi_1$, the LC50 and RC50 curves represent the cases where the BR of $\chi_{1}\rightarrow t\ell$ decay mode is $50$\%. Although they are not realized in the LC and RC scenarios, such a situation is possible if there are other decay modes of $\chi_1$ (which play no role in our analysis beyond modifying the BR). Hence, we show these plots to give some estimates of how the significance would vary with the BR.
\item 
For comparison, we also show the expected significance obtained with only the pair production events for the $50\%$ and $100\%$ BR cases.
For instance, for $100\%$ BR in the $\chi_1\to t\ell$ mode, the HL-LHC (3 ab$^{-1}$) discovery mass reach (i.e., $\mc Z=5\sg$) with only pair production is about $2.05~(2.35)$  TeV for $\kp=0~(\kp=1)$.
\item When the LC coupling is unity, the discovery reach goes up to $2.35~(2.50)$ TeV once the single production processes are included. However, in the RC scenario the improvement is  minor.
This happens because  $\sg\lt(pp\to \chi_1\ell j\rt)$ is larger in the LC scenario than that in the RC scenario. 
\item
Unlike the scalar case, there is no interference among the different signal diagrams, and hence the signal significances in the  RLCSS or RLCOS benchmarks are the same as that in the RC scenario.  
\item 
In Figs.~\ref{fig:xset02} and ~\ref{fig:xset12} we observe that the maximum reach for $\chi_5$ comes from the combined LC scenario. The values are $2.35$ and $2.50$ TeV for $\kappa =0$ and $\kappa =1$, respectively. There is a suppression in the RC channel for a similar reason as for $\chi_{1}$, a $\chi_{5}$ LQ also couples to a right chiral top.
\end{itemize}
In Table~\ref{tab:sig} we collect all of the numbers for $\mc Z=2\sg$, $3\sg$, and $5\sg$.

Since we can parametrize the combined signal cross section for any $M_\chi$ as
\be
\sg_{\rm signal} \approx \sg_{\rm pair}(M_\chi) + \lm^2 \sg_{\rm single}(\lm=1,M_\chi),
\ee
the combined signal cross section increases with $\lm$ for any fixed $M_\chi$. 
By recasting the figures shown in Fig.~\ref{fig:xset}, which are for $\lm=1$, we can obtain the reach in the $\lm$-$M_\chi$ plane, as we show in Figs.~\ref{fig:xlamv1z5r} 
and \ref{fig:xlamv1z2s}. We show the $5\sg$ discovery curves in Fig.~\ref{fig:xlamv1z5r} while the $2\sg$ exclusion curves are displayed in Fig.~\ref{fig:xlamv1z2s}.
These plots show the lowest value of $\lm$ required to observe the vLQ signal for a varying $M_\chi$ with $5\sg$ confidence level for discovery. For the exclusion plots, all points above the
curves can be excluded at the $95\%$ confidence level at the HL-LHC.

\section{Summary and conclusions}\label{sec:End}

\noindent
Usually, in the direct LQ searches, it is assumed that LQs only couple to quarks and leptons of the same generation. Collider signatures of TeV-scale LQs with large cross-generational couplings, motivated by the persistent flavor anomalies, are completely different than what is considered in the usual LQ searches at the LHC. It is then important to explore these possibilities in detail. 
In a previous paper~\cite{Chandak:2019iwj}, we investigated the HL-LHC prospects of all scalar LQ models within the Buchm\"uller-R\"uckl-Wyler classifications~\cite{Buchmuller:1986zs} that would produce  \emph{boosted-$t$ $+$ high-$p_{\rm T}$-$\ell$} signatures at the LHC.
In this follow-up paper, we investigated the case for the vector LQs with the same signature. The vLQs that decay to a top-quark can have three possible electric charges, $\pm 1/3$, $\pm 2/3$, and $\pm 5/3$. Among these, our primary focus was on the charge $\pm 1/3$, $\pm 5/3$ vLQs that can decay to a top quark and an electron or a  muon as a unique top-lepton
resonance system would appear from the decays of these LQs. 

In this paper, we introduced some simple phenomenological Lagrangians. These simple models can cover the relevant parameter spaces of the full models described in Refs.~\cite{Buchmuller:1986zs,Dorsner:2016wpm}. In this simplified framework, we studied the pair and single production channels of vector LQs at the LHC. Pair production of the vLQs produces final states with two boosted top quarks and two high-$p_T$ leptons and determines the LHC discovery reach in the low-mass region. On the other hand, the single production processes produce final states with at least one boosted top quark and two high-$p_T$ leptons. We observed two interesting points about single production. 1) Despite considering vLQ couplings with only third-generation quarks, we see that the single production cross sections are not necessarily very small, provided, of course, the new couplings controlling them are not negligible. 2) Like the pair production, some single production processes can also depend on the parameter $\kappa$ that appears in the gluon-vector LQ coupling. In some scenarios, for order-one new coupling(s), the single production would control the LHC reach in the high-mass region.

We adopted a search strategy of selecting events with at least one boosted hadronic top quark and exactly two high-$p_ {\rm T}$ leptons of the same flavor and opposite sign. This combines events from the pair and single productions and, therefore, enhances the discovery reach by about $300$ GeV from the usual pair production searches at the LHC. Our results show that charge $1/3$ and $5/3$ vector LQs can be probed up to $2.35$ ($2.50$) TeV for $100\%$ branching ratio in the $t\ell$ decay mode for $\kp=0$ $(\kp=1)$ and order-one new couplings at the $14$ TeV LHC with $3$ ab$^{-1}$ of integrated luminosity with $5\sg$ significance. Alternately, in the absence of their discoveries, they can be excluded up to $2.65$ $(2.75)$ TeV at the $95\%$ confidence limit. Since the single production cross sections scale as $\lm^2$, we also showed how the discovery/exclusion reach would vary with $\lm$ within its perturbative domain.

Comparing with the results in Ref.~\cite{Chandak:2019iwj}, we saw that in general it may be possible for the LHC to discover/exclude heavier vLQs than sLQs for comparable parameters. For example, for $\lm=1$ the $5\sg$ discovery reach for $\chi_5$ goes up to $2.35$ TeV ($\kp=0$) in the LC scenario. This is higher than the $1.75$ TeV reach in the LC scenario for $\phi_5$ (or even $1.95$ TeV in the RC scenario). Similar observations can be made for other LQs/scenarios too. Of course, in the case of vLQs, the additional parameter $\kp$ can enhance the reaches even more: for $\chi_5$ in the LC scenario it increases to $2.5$ TeV. There is no such parameter in the case of sLQs: in this regard, the sLQ pair production process is more model independent (i.e. QCD driven) than that for vLQs. Also, for some sLQs, the relative signs between new physics couplings are important as they affect the single production cross sections through interference among different signal diagrams. This happens for $\phi_1$ whose single production cross sections in the LCSS and LCOS scenarios differ significantly. However, there is no such interference for vLQs; for example, the $\chi_1$ single production cross section is the same in the RLCSS and RLCOS scenarios. Of course, for both sLQs as well as vLQs, single production processes do play an important role in determining the LHC discovery/exclusion reaches even though we considered only third-generation quarks coupling with the LQs. It would be interesting to investigate ways to distinguish sLQs and vLQs with such similar signatures at the LHC. Finally, we point out that in these two papers we have presented simplified models for all possible LQs that couple with the top quark and leptons. These simplified modes are suitable for bottom-up/experimental studies and can be easily mapped to the full models.

\acknowledgments 
\noindent 
A. B. and S.M. acknowledge financial support from the Science and
Engineering Research Board (SERB), DST, India under grant number ECR/2017/000517.

\bibliography{Leptoquark}{}
\bibliographystyle{JHEPCust}

\end{document}